\documentclass[
	aps,
	showpacs,
	showkeys,
	twocolumn,
	altaffilletter,
	nolongbibliography,
	numerical,
	flushbottom,
	secnumarabic,
	pra,
	superscriptaddress,
	floatfix,
	10pt
]{revtex4-1}

\usepackage{lipsum}

\usepackage[dvipsnames]{xcolor}

\usepackage{bm}

\usepackage{graphicx}

\usepackage{upgreek}

\usepackage{natbib}

\usepackage{braket}

\usepackage{physics}
\usepackage{newtxtext, newtxmath}

\usepackage{siunitx}

\usepackage[%
	breaklinks,%
	pdftex,%
	hyperfootnotes=true,%
	pdfpagelabels,%
	bookmarks,%
	pageanchor,%
]{hyperref}

\hypersetup{%
	colorlinks=true, linktocpage=true, pdfstartpage=1, pdfstartview=FitH, pdfborder={0 0 0},%
	breaklinks=true, pdfpagemode=UseNone, pageanchor=true, pdfpagemode=UseOutlines,%
	plainpages=false, bookmarksnumbered, bookmarksopen=true, bookmarksopenlevel=1,%
	hypertexnames=true, pdfhighlight=/O,
	urlcolor=Red!50!Sepia, linkcolor=NavyBlue, citecolor=RoyalBlue, 
}

\graphicspath{{Pictures/}}

\newcommand{\ie}{i.\,e.}

\newcommand{\eg}{e.\,g.}
\newcommand{\ped}[1]{_\text{#1}}
\newcommand{\api}[1]{^\text{#1}}

\newcommand{\ham}{H}

\newcommand{\rng}[2]{\ensuremath{[#1, #2]}}
\renewcommand{\epsilon}{\varepsilon}

\newcommand{\beq}{\begin{equation}}
\newcommand{\eneq}{\end{equation}}

\DeclareMathOperator{\eu}{e}
\DeclareMathOperator{\iu}{i}

\begin{document}

\author{L.~M.~Cangemi}
\affiliation{Dipartimento di Fisica ``E.~Pancini'', Universit\`a di Napoli ``Federico II'', Complesso di Monte S.~Angelo, via Cinthia, 80126 Napoli, Italy}
\email{lorismaria.cangemi@unina.it}
\affiliation{CNR-SPIN, c/o Complesso di Monte S. Angelo, via Cinthia - 80126 - Napoli, Italy}
\author{V.~Cataudella}
\affiliation{Dipartimento di Fisica ``E.~Pancini'', Universit\`a di Napoli ``Federico II'', Complesso di Monte S.~Angelo, via Cinthia, 80126 Napoli, Italy}
\affiliation{CNR-SPIN, c/o Complesso di Monte S. Angelo, via Cinthia - 80126 - Napoli, Italy}
\author{G.~Benenti}
\affiliation{Center for Nonlinear and Complex Systems, Dipartimento di Scienza e Alta Tecnologia, Universit\`a degli Studi dell'Insubria, via Valleggio 11, 22100 Como, Italy} 
\affiliation{NEST, Istituto Nanoscienze-CNR, I-56126 Pisa, Italy}
\affiliation{Istituto Nazionale di Fisica Nucleare, Sezione di Milano, via Celoria 16, 20133 Milano, Italy}
\author{M.~Sassetti}
\affiliation{Dipartimento di Fisica, Universit\`a di Genova, Via Dodecaneso 33, 16146 Genova, Italy} 
\affiliation{CNR-SPIN,  Via  Dodecaneso  33,  16146  Genova, Italy}
\author{G.~De Filippis}
\affiliation{Dipartimento di Fisica ``E.~Pancini'', Universit\`a di Napoli ``Federico II'', Complesso di Monte S.~Angelo, via Cinthia, 80126 Napoli, Italy}
\affiliation{CNR-SPIN, c/o Complesso di Monte S. Angelo, via Cinthia - 80126 - Napoli, Italy}

\title{Violation of TUR in a periodically driven work-to-work converter from weak to strong dissipation}

\date{\today}


\begin{abstract}
We study a model of isothermal steady-state work-to-work converter, where a single quantum two-level system (TLS) driven by time-dependent periodic external fields acts as the working medium and is permanently put in contact with a thermal reservoir at fixed temperature $T$.~By combining Short-Iterative Lanczos (SIL) method and analytic approaches, we study the converter performance in the linear response regime and in a wide range of driving frequencies, from weak to strong dissipation.~We show that for our ideal quantum machine several parameter ranges exist where a violation of Thermodynamics Uncertainty Relations (TUR) occurs.~We find the violation to depend on the driving frequency and on the dissipation strength, and we trace it back to the degree of coherence of the quantum converter.~We eventually discuss the influence of other possible sources of violation, such as non-Markovian effects during the converter dynamics.
\end{abstract}

\maketitle

\section{Introduction}\label{sec:intro}

    Energy conversion at the microscopic scale poses great experimental and theoretical challenges. In the classical setting, models of stochastic cyclic heat engines are among the prototypical systems of interest \cite{Schmiedl_2007,Holubec_2014}.~Experimentally, a heat engine with a single optically-trapped, micrometre-sized particle acting as the working medium, subject to periodically-driven forces and put in contact with two thermal reservoirs was realized \cite{Blickle:StocHeatEng}.~More recently, the experimental realization of Brownian Carnot cycles has been achieved \cite{Martinez:BrownEng}.    
    
    As the driving forces vary on timescales smaller than the thermal relaxation time, in these engines the thermal fluctuations arising from stochastic forces acting on the working medium cannot be neglected; as a consequence, the concepts of classical thermodynamics,~\ie~heat, work and entropy production, need to be generalized to microscopic nonequilibrium regime, and the fundamental limits set to heat-to-work conversion have to be reconsidered.~The theoretical framework of Stochastic Thermodynamics \cite{Seifert_2012:Review,Seifert:annual,Sekimoto:StochEne}, based on universal nonequilibrium fluctuation theorems \cite{Bochkov:General,Jarzynski:Work,Crooks:WorkFluct}, has been employed to model heat, work and entropy production as stochastic quantities and to describe energy conversion in these microscopic engines.~The search for optimal working performance of these machines is relevant in the so-called field of finite-time thermodynamics \cite{VanDenBroeck:thermomaxpow,Esposito:MaxPowEff,Andresen:finitetime}.    
    
    Steady-state thermal machines,~\eg~thermoelectric devices 
    coupled to time-independent reservoirs \cite{BENENTI20171}, belong to the class of autonomous thermal machines.~In this context, several interesting results have been derived,~\eg~in the absence of time-reversal (TR) symmetry due to the presence of a magnetic field, the second law of thermodynamics by itself does not forbid the possibility of achieving the Carnot efficiency at finite power \cite{Benenti:TR}.~Such possibility was denied by subsequent studies using more specific assumptions \cite{Balach:eff,Brandner:StrongBound,Brandner2:multiterm13,Brandner:Bound15,Yamamoto:Effbound,Shiraishi:untradeoff} or symmetry considerations for the kinetic (Onsager) coefficients \cite{luo2019onsager}.~In particular, for classical heat engines whose interactions with heat baths can be described as Markov processes, it was proved \cite{Shiraishi:untradeoff} that the mean power $P$ has an upper bound, $P\le A(\eta_{\rm{\scriptscriptstyle C}}-\eta)$, where $\eta$ is the engine efficiency, upper bounded by the Carnot efficiency $\eta_{\rm{\scriptscriptstyle C}}$, and $A$ is a system-specific amplitude.~While at first sight such bound implies, for a broad class of systems, that $P\to 0$ when $\eta\to\eta_{\rm{\scriptscriptstyle C}}$,~\ie~the Carnot bound can be achieved only in the infinite working time limit, thus producing zero output power, one cannot exclude that the prefactor $A$ diverges when approaching the Carnot efficency \cite{Allahver:Bounds,Campisi:CriticalHeatEng}, for instance at the verge of a quantum phase transition \cite{Campisi:CriticalHeatEng}.~The divergence of fluctuations when approaching a phase transition suggest that one should consider a third quantity when characterizing the performance of a heat engine, besides power and efficiency, that is, power fluctuations \cite{Pietzonka:Tradeoff}. 
   
    Indeed, quite recently, a set of tradeoff relations have also been estabilished in the classical domain, linking the entropy production to the power output and power fluctuations,~\ie~the so-called Thermodynamics Uncertainty Relations (TUR) \cite{Pietzonka:Tradeoff,Horowitz:TUR}.~TUR rule the tradeoff between entropy production and the output power relative fluctuations,~\ie~the precision of the machine, so that working machines operating at near-to-zero entropy production cannot be achieved without a divergence in the relative output power fluctuations.~Quite recently, a generalization of TUR has been provided for periodically-driven systems, and operationally-accessible bounds to entropy production,~\ie~written in terms of quantities directly accessible to experiment \cite{Barato_2018:bound,Koyuk:BoundPer} have been proposed.       
      
    So far, the implications of quantum mechanics on the mechanism of heat-to-work conversion have not been completely understood \cite{Alicki:introQT}.~Fluctuation relations have been generalized to the quantum domain \cite{Esposito:Noneq1,Campisi:Noneq2}, so that a theoretical description of quantum fluctuations of heat and work has been provided.~However, the work spent on the systems by means of a fixed time protocol in principle requires two projective measurement, at the start and at the end of the protocol respectively.~It is thus difficult to probe work statistics \cite{Talkner:WorknotObs}, and several measurement strategies have been devised to solve the issue \cite{Batalhao:Flucttheo,Roncaglia:work}.~A further fundamental and delicate problem is the same definition of work, heat and entropy production in the presence of strong correlations with the thermal reservoirs \cite{Alipour:Correlations,Bera:thermocorr}.
    
    However, in the last decades different models of quantum heat engines and refrigerators have been devised, where the working medium can consist of a single driven TLS \cite{AllahVer:WorkSBM, Geva:tlseng1,Alicki:Minimal,Uzdin:equivalence}, couples of harmonic oscillators \cite{Lev:RefrigHarmonicosc,Insinga:HarmEng}, pairs of qubits subject to unitary gates \cite{CampisiPekolaFazio:quantheng}, quantum dots, autonomous motors made of Brownian particles \cite{Drewsen:DuetBrown}, as well as many body systems near criticality \cite{Campisi:CriticalHeatEng}.~Otto cycles, Carnot cycles, both in the adiabatic and finite-time configuration have been considered, where the working medium undergoes a finite number of strokes \cite{Uzdin:equivalence}, and it is put in contact with two thermal reservoirs.~In these works, different interesting results for the efficiency at maximum power have been reported \cite{Allahver:Bounds,CampisiPekolaFazio:quantheng}, 
    while the quantum engines performance are found to severely depend on the speed of the protocol and on the bath spectral properties \cite{Cavina:Slowthermo,Erdman:MaxPow}.~Shortcuts to adiabaticity \cite{DelCampo:MoreBang,Abah:STA1,Cakboh:STA2,hartmann:STAmanybody} have also been considered, which can reduce the effect of friction and provide noticeable improvements in the performance of cyclic heat engines.    
     
    Despite all these efforts, it remains unclear whether or not the quantum nature of the working medium can provide a relevant enhancement in the efficiency of heat to work conversion.~Moreover, the experimental realizations of these devices remain limited \cite{Pekola,Robnagel,SpinHeateng,Roche:Harvest,Thierschmann:Harvest,Josefsson:qdoteng,Jaliel:Harvest}.  
    
    A noticeable part of the works in the literature model the coupling with the thermal reservoirs by making use of Quantum Master Equations (QME), thus limiting the analysis to the weak coupling regime, while the mechanism of heat to work conversion in the strong coupling regime has received much less attention \cite{Nazir:Strongcoup1,PerarnauLobet:strongcoup2,Katz:strongcoup3,Gelbwaser:strongcoup4,Wiedmann2020:outofeq}.~In the case of quantum engines, recent works reported possible violations of the TUR occurring both in the steady \cite{Ptaszy:vioTUR1,Agarwalla:vioTUR2} and in the presence of driven cases \cite{Carrega:wtow}.~Moreover, for periodically-driven quantum engines, the actual validity of the TUR remains highly controversial \cite{Guarnieri:Thermoprecision}.    

    Below, we consider a simple model of periodically driven, isothermal machine, where the working medium consists of a single TLS, driven by two external periodic fields of fixed amplitudes and permanently put in contact with a thermal bath.~Similar machines, which have been recently discussed in the classical setting using linear irreversible thermodynamics \cite{Proesman:powereff}, and in the quantum setting employing models of adiabatic quantum pumps \cite{Ludovico:Adiabresp,Juergens:thermoel,Bustos:Thermo}, act as prototypes of work-to-work converters. In these machines, a given amount of 
    work provided in the input channel is converted to the output channel with fixed efficiency.~Employing a standard definition of work \cite{Wiedmann2020:outofeq}, we numerically simulate the dynamics of this engine for different values of the model parameters.~Our numerical approach, based on Short Iterative Lanczos (SIL) method, allows us to provide a reliable description of the system dynamics from weak to moderately strong dissipation, in the low-temperature regime, beyond the capabilities of conventional QME treatments and with fairly no limitations on the value of the driving frequency and fields amplitudes.~Restricting to linear response regime, we compute the efficiency, the output power and fluctuations as functions of the model parameters,~\ie~driving frequency and dissipation strength.~Combining our numerical approach with exact analytic results, we show that a violation of the TUR for periodically driven systems can occur in a wide range of model parameters, so that different working regimes exist in which the quantum converter may achieve a better tradeoff between entropy production and output power fluctuations. 
    
    The work is organized as follows: in Sec.~\ref{sec:Chap3sec1}, we describe our model of simple work-to-work converter.~In Sec.~\ref{sec:balancetur} we discuss the main quantities of interest to measure the converter performance and how they enter into the different formulations of TUR.~In Sec.~\ref{sec:linearregime} we describe how to simulate the converter performance restricting to linear response regime.~In Sec.~\ref{sec:violations} we report our numerical results for the converter performance and TUR violations in the linear response.~We also discuss interesting parameter regimes where the linear response can be computed analytically.  
   
    \section{Setup of the converter}\label{sec:Chap3sec1}
\begin{figure}[h]
    	\centering
    	\includegraphics[scale=1.0]{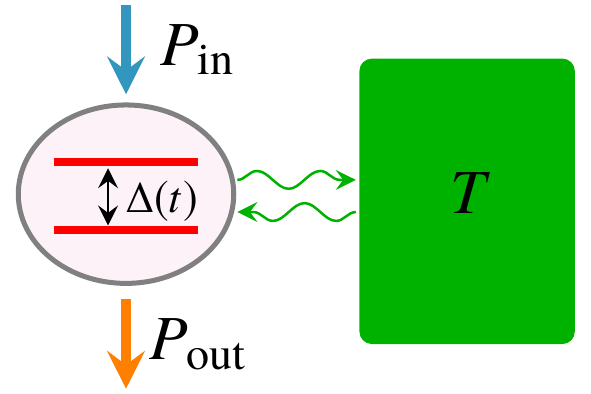}
    	\caption{Schematic diagram of the converter: the TLS bias is periodically modulated in time, while the TLS is in contact with a thermal reservoir at fixed temperature $T$.}
    	\label{fig:converter}
    \end{figure}

    We model our system with a TLS in the presence of two external time-periodic driving fields of fixed amplitudes, phase difference and frequency $\omega$.~The TLS is in contact with a heat bath at fixed temperature $T$, as sketched in Fig.~\ref{fig:converter}.~The Hamiltonian of the whole system can be written as
    \begin{equation}\label{eq:TotalHamchap3}
    H(t)= H\ped{S}(t) + H\ped{B} + H\ped{SB}, 
    \end{equation}
    where $H\ped{S}(t)$, $H\ped{B}$ are respectively the free Hamiltonian of the system and bath and $H\ped{SB}$ is the interaction energy between the TLS and the bath.  
   
    We choose the following form for the TLS Hamiltonian 
    \begin{equation}\label{eq:QubitHamChap3}
    H\ped{S}(t)=  -\frac{1}{2}(\epsilon_1(t)+\epsilon_2(t) )\sigma_z -\frac{\Delta}{2} \sigma_x.
    \end{equation}      
    Here $\Delta$ is the tunnelling element, and $\epsilon_1(t),\epsilon_2(t)$ are two external periodic driving fields,~\ie~$\epsilon\ped{i}(t)=\epsilon\ped{i}(t+\small{\mathcal{T}})\mbox{, } i=1,2$, oscillating with period $\mathcal{T}$, which modulate the levels asymmetry.~Here and in the following, we set $\hbar=1$.~While
    there is no limitation on the detailed form of the driving fields, we fix them as follows: $\epsilon_1(t)=\epsilon_1 \sin\omega t\mbox{, }
    \epsilon_2(t)=\epsilon_2 \cos(n\omega t-\varphi)$, where $\epsilon_1,\epsilon_2$ are the driving field amplitudes, $\omega=2\pi/\mathcal{T}$ is the driving frequency and $\varphi$ is the phase difference.~We choose as a set of basis states for the TLS the eigenstates of $\sigma_z$ operator,~\ie~$\sigma_z\ket{\pm}=\pm\ket{\pm}$.~The bath is modeled with a set of bosonic oscillators of frequency $\omega_k$. As a consequence, adopting the formalism of bosonic creation (annihilation) operators $b^{\dagger}_k (b^{\phantom{\dagger}}_k) $, the Hamiltonian of the free bath $H\ped{B}$ can be written as  
    \begin{equation}\label{eq:BathHamChap3}                
    H\ped{B}=\sum_k\omega_k b^{\dagger}_k b_k. 
    \end{equation} 
    The interaction Hamiltonian $H\ped{SB}$ can be modeled as customary in the Spin Boson Model (SBM) literature \cite{Leggett,SassettiWeiss:SpinBoson,Sassetti:Univ,LeHurNRG1,Grifoni:drivenquantum,Grifoni:Coop}, where the TLS couples linearly to the bath degrees of freedom along the $z$ direction,  
    \begin{equation}\label{eq:IntHamChap3}
    H\ped{SB}=\frac{1}{2}\sigma\ped{z}\sum_k \lambda_k (b^{\dagger}_k + b_k ).
    \end{equation}    
    In Eq.~\eqref{eq:IntHamChap3}, $\lambda_k$ is the coupling strength with the k-th oscillator.~The properties of the bath are characterized by the spectral density function \cite{Leggett}
    \begin{equation}\label{eq:SpectDensityChap3}
    J(\omega)= \sum_k \lambda_k^2 \delta(\omega-\omega_k)= 2 \alpha \frac{\omega^s}{\omega^{s-1}_c}\eu^{-\frac{\omega}{\omega\ped{c}}}.
    \end{equation}  
    It can be written as a sum over discrete frequencies of the bath modes, ranging from $0$ up to a cutoff frequency $\omega\ped{c}$ and, in the continuum limit, it can be expressed as the right-hand side of \eqref{eq:SpectDensityChap3}.~The adimensional parameter $\alpha$ is a measure of the strength of the dissipation, which is a consequence of the linear coupling of the system to the whole set of oscillators.~Different kinds of dissipation can be described by means of the parameter $s$ \cite{Weiss:open-quantum2,Leggett,bulla:nrg2}: Ohmic, sub-Ohmic and super-Ohmic dissipation cases correspond to $s=1$, $s<1$ and $s>1$ respectively.        
    
    In what follows, we focus on the Ohmic regime.~In the absence of driving fields, the dynamics of the TLS can be described by means of approximate analytic approaches.~Starting from a factorized state of the TLS and the bath, at $T=0$ the longitudinal magnetization $\ev{\sigma_z (t)}$ undergoes underdamped oscillations in time with frequency
    \begin{equation}\label{eq:Deltaeff}
    \Delta\ped{eff}=\Delta\ped{r}\left[\Gamma(1-2\alpha) \cos\pi \alpha\right]^{1/(2(1-\alpha))},
    \end{equation}
    where $\Delta\ped{r}=\Delta(\Delta/\omega\ped{c})^{\alpha/(1-\alpha)}$ is the renormalized gap of the TLS and $\Gamma(z)$ is the Gamma function.~As a consequence, with increasing dissipation strength the TLS dynamics becomes progressively incoherent.
    
    On the other hand, in the case of Eq.~\eqref{eq:TotalHamchap3}, the TLS,~\ie~the working medium, is driven out of equilibrium by means of the two external fields, while the permanent contact with the bath induces dissipation and decoherence.~Although Floquet theory provides a satisfactory description of the nonequilibrum dynamics of periodically-driven systems in the absence of system-bath interactions, the physical description of the open system dynamics beyond conventional Born-Markov approximation is still incomplete \cite{Weiss:open-quantum2,Grifoni:drivenquantum,Magazzu:Floquetasy,Hartmann:floquetmaps,Restrepo_2018}. 
    For this class of driven open quantum systems, at long times a nonequilibrium stationary state is expected, where the reduced density matrix of the TLS undergoes periodic time evolution with period $\mathcal{T}$.~In this regime, the dynamics of energy exchange shows that the expectation values of the different operators in Eq.~\eqref{eq:TotalHamchap3},~\ie~$\ev{H\ped{S}(t)},\ev{H\ped{B}},\ev{H\ped{SB}}$, exhibit a periodic evolution with period equal to $\mathcal{T}$.~However, the constant time-averaged power injected by the external drive into the system is entirely drained by the bath,~\ie~the powers drained by the TLS and the interaction channel SB average to zero over a period $\mathcal{T}$ \cite{SassettiWeiss:energyexchange}.              
    
    Several parameter ranges of the model in Eq.~\eqref{eq:TotalHamchap3} exist where the two fields behave as input and output channels,~\ie~the mean powers of the two different driving fields take opposite signs and the whole system acts as a work-to-work converter.~Given the discrete nature of the energy levels of the working medium, the converter could be experimentally realized by employing superconducting circuits,~\eg~qubits driven out-of-equilibrium by means of two electromagnetic signals \cite{IngoldNazarov:tunnel,XiangNori:hybrid,Krantz:engqubits}.~Here the environment describes the effects of noise introduced by external circuitry~\cite{Girvin:cQED,Devoret1997QuantumFI}.
    
    For sufficiently small amplitudes of the driving fields with respect to the driving frequency $\omega$ and the tunnelling element $\Delta$, a description based on linear irreversible thermodynamics \cite{BENENTI20171,Proesman:powereff} can thus be employed: here the converter can be studied as a two terminal steady-state device, where the ratios of the expectation values of input and output powers to the corresponding field amplitudes $(\epsilon_1,\epsilon_2)$ play the role of the currents, while the field amplitudes act as the thermodynamic forces.
   
    A similar approach has been followed in \cite{Carrega:wtow}, where in place of a single TLS a quantum Brownian particle in a tight-binding lattice has been considered: in the weak tunnelling regime, it has been shown that violations of static TUR can occur.~However, a theoretical study of the TUR for periodically driven systems, employing a model of quantum work-to-work converter as in Eq.~\eqref{eq:TotalHamchap3}, beyond weak-tunnelling approximation and for strong system-bath coupling has not been reported.     
   
    In the subsequent sections, we study the nonequilibrium properties of the converter described in Eq.~\eqref{eq:TotalHamchap3}: we first find regions in the parameter space where work-to-work conversion occurs; then, restricting to linear response regime, we find evidence of systematic violations not only of the static \cite{Pietzonka:Tradeoff}, but also of the TUR for periodically driven systems \cite{Koyuk:BoundPer} (dynamic TUR from now on), which show up for weak dissipation and in the low temperature regime. 
    \section{Energy balance and TUR}\label{sec:balancetur}
    The dynamics of system in Eq.~\eqref{eq:TotalHamchap3} is described by means of the total density matrix $\rho(t)$, which undergoes a unitary evolution obeying Von Neumann equation of motion
   \begin{equation}\label{eq:VonNeumann2}
   \frac{\mathrm{d}}{\mathrm{d} t}\rho(t)=-i\comm{H(t)}{\rho}.
   \end{equation}   
    Choosing the initial state of the total density matrix $\rho(t\ped{0})$ (see Sec.~\ref{subsec:convdyn}), Eq.~\eqref{eq:VonNeumann2} allows us in principle to compute the density matrix of the whole system at any subsequent time $t$.~Thus, the reduced density matrix of the TLS can be computed as 
   \begin{equation}\label{eq:reduceddens2}
   \rho\ped{S}(t)=\tr\ped{B}\rho(t),
   \end{equation}
   where the partial trace is taken over the bath degrees of freedom.~From the knowledge of $\rho(t)$, the expectation value of the total energy of the system can be computed as follows
   \begin{equation}
   \ev{H(t)}=\tr\left[H(t) \rho (t) \right]. 
   \end{equation}
   Due to the external driving fields, the expectation value of the total energy of the system changes in time.~From Eq.~\eqref{eq:TotalHamchap3}, it follows
   \begin{equation}
   \frac{\mathrm{d}}{\mathrm{d} t}\ev{H(t)}=\tr\left[ \frac{\partial H\ped{S}(t) }{\partial t} \rho(t) \right]. 
   \end{equation}
   We can thus define the mean power linked to each driving field as follows
   \begin{equation}\label{eq:powerst}
   \ev{P_i (t)}=-\frac{1}{2} \dot{\epsilon}_i (t)\ev{\sigma\ped{z}(t)},  
   \end{equation}
   where $i=1,2$, so that $\ev{P\ped{1}(t)}+\ev{P\ped{2}(t)}=\mathrm{d}\ev{H(t)}/\mathrm{d}t$. Eq.~\eqref{eq:powerst} is also equal to the average work per unit time linked to the channel $i$. 
 
   In general, the dynamics of $\rho(t)$ is nontrivial, due to the combined effects of the driving fields and dissipation.~However, for sufficiently long times a nonequilibrium time-periodic steady state is reached, where the state of the system evolves periodically in time with period $\mathcal{T}$.~Hence, at long times the time-averaged expectation values of the mean powers are relevant,~\ie  
   \begin{equation}\label{eq:meanpowers}
   \overline{P}_i=\frac{1}{\mathcal{T}}\int\limits_t^{t+\mathcal{T}}\ev{P_i (t^{\prime})}\mathrm{d}t^{\prime}. 
   \end{equation}
   As it follows from analytical results \cite{SassettiWeiss:energyexchange,Wiedmann2020:outofeq}, at long times the total power injected into the system is drained by the bath,~\ie~the time derivatives of the expectation values $\ev{H\ped{S}(t)},\ev{H\ped{SB}}$ average to zero over a period $\mathcal{T}$.~It can be readily verified by computing numerically the time evolution of the former expectation values (see Sec.\ref{subsec:conveff}).~This property allows to unambiguosly define the heat exchanged with the bath per unit time,~\ie~$\dot{\mathcal{W}}\ped{B}$ as the energy flowing into the bath per unit time, which reads        
   \begin{equation}\label{eq:heatpertime}
   \dot{\mathcal{W}}\ped{B}=- \tr\left[H\ped{B}\dot{\rho}(t)\right]= i \tr \left[\comm{H\ped{B}}{H\ped{SB}}\rho(t)\right].    
   \end{equation}
   Integrating Eq.~\eqref{eq:heatpertime} over a period, from Eq.~\eqref{eq:meanpowers} we find 
   \begin{equation}\label{eq:firstlaw}
   \mathcal{W}\ped{B}= -\int\limits_{t}^{t+\mathcal{T}}\tr\left[H\ped{B}\dot{\rho}(t)\right]\mathrm{d} t=\mathcal{T}(\overline{P}_1+\overline{P}_2). 
   \end{equation}
   Eq.~\eqref{eq:firstlaw} fixes the energy balance of our machine in the nonequilibrium periodic steady-state.~The system operates as a work-to-work converter if the average powers $\overline{P}_1\mbox{, }\overline{P}_2$ take opposite signs,~\ie~a part of the work spent per unit time in a given channel is converted in the other channel.~We are interested in the conversion efficiency, which can be written as
   \begin{equation}\label{eq:eff}
   \eta = \frac{\abs*{P\ped{out}}}{P\ped{in}},
   \end{equation}
   where we conventionally take as the output channel the one which brings negative power,~\ie~$P\ped{out}=\overline{P}_1(\overline{P}_2)$, $P\ped{in}=\overline{P}_2(\overline{P}_1)\mbox{ if }\overline{P}_1(\overline{P}_2)<0\mbox{ and }\overline{P}_2(\overline{P}_1)>0$.~In heat-to-work conversion, for a system connected to two baths at different temperatures, $\eta$ is upper bounded by the Carnot efficiency.~In the present isothermal work-to-work conversion the upper bound for efficiency is $\eta=1$.~Achieving a conversion efficiency close to $1$ means that only a small amount of the power spent in input is dissipated into the bath.
  
   At long times $t$,~\ie~when the nonequilibrium stationary state has been reached, the output power fluctuations of our converter can be computed as follows
   \begin{equation}\label{eq:fluct}
   D_i (t)=\int\limits_0^{+\infty}\mathrm{d}\tau (\ev{\delta P_i (t) \delta P_i (t-\tau)}+\ev{\delta P_i (t-\tau) \delta P_i (t)}),  
   \end{equation}
   where $P\ped{i}(t)$ is the power operator and $\delta P_i (t)=P_i (t)-\ev{P_i (t)}$.~The brackets 
   denote the quantum mechanical expectation value to be computed using the whole system + bath density matrix at time $t$.~Inserting Eq.~\eqref{eq:powerst} into Eq.~\eqref{eq:fluct} and taking the time average over a period, the time-averaged power fluctuations can be expressed in terms of the two-time correlation function $\mathcal{B}(t,t-\tau)=\Re\{\ev{\sigma\ped{z}(t)\sigma\ped{z}(t-\tau)}\}-\ev{ \sigma\ped{z}(t)}\ev{\sigma\ped{z}(t-\tau)}$ as follows    
   \begin{equation}\label{eq:averagecorr}
   \overline{D}_i=\frac{1}{2\mathcal{T}}\int\limits_0^\mathcal{T} \mathrm{d}t\dot{\epsilon_i}(t)\int\limits_0^{+\infty}\mathrm{d}\tau \dot{\epsilon_i}(t-\tau)\mathcal{B}(t,t-\tau).  
   \end{equation}  
   From Eq.~\eqref{eq:meanpowers},~\eqref{eq:eff},~\eqref{eq:averagecorr}, it is possible to investigate the tradeoff between output power, entropy production and output power fluctuations for the work-to-work converter in the nonequilibrium steady-state regime. 
   
   As anticipated in the introduction, in the classical thermodynamics setting and for static external fields it has been shown that the average currents in microscopic steady-state devices are linked to their mean fluctuations and to entropy production $\sigma$ by means of TUR. In the case of our converter, TUR reads \cite{Pietzonka:Tradeoff,Horowitz:TUR}          
   \begin{equation}\label{eq:tur}
   \mathcal{Q}=\sigma \frac{D\ped{out}}{P^2\ped{out}} \geq 2.  
   \end{equation}
   Here it is worth rewriting the left hand side of Eq.~\eqref{eq:tur},~\ie~our tradeoff parameter, in terms of the conversion efficiency in Eq.~\eqref{eq:eff} as follows
   \begin{equation}\label{eq:tureff}
   \mathcal{Q}=\sigma \frac{D\ped{out}}{P^2\ped{out}}=\beta \abs{P\ped{out}} \left(\frac{1}{\eta}- 1\right)\Sigma^2\ped{out} \geq 2,  
   \end{equation}
   where $\sigma=\frac{1}{T} (P_{\text{in}}- \abs{P_{\text{out}}})$, $\beta=1/T$ is the inverse temperature and $\Sigma\ped{out}=\sqrt{D\ped{out}/P^2\ped{out}}$ is the relative power uncertainty.~Eq.~\eqref{eq:tureff} sets a lower bound to the product of output power fluctuations and entropy production at fixed output power, so that in the TR symmetric, reversible operating regime the divergence of relative fluctuations follows.~Several works have reported violations of TUR in the quantum realm \cite{Liu:TURViol,Ptaszy:vioTUR1,Agarwalla:vioTUR2,Carrega:wtow}, and a possible explanation has been proposed in \cite{Guarnieri:Thermoprecision}, pointing towards a smaller lower bound set by quantum mechanics with respect to the classical case.     
   
   In the case of periodically-driven nonequilibrium engines, TUR have been recently generalized \cite{Koyuk:BoundPer} as follows
   \begin{equation}\label{eq:turperiodic}
   \sigma(\omega) \frac{D\ped{out}(\omega)}{P^2\ped{out}(\omega)} \geq  \mathcal{V}_{\rm \scriptscriptstyle{TUR}}(\omega),
   \end{equation}  
   where 
   \begin{equation}\label{eq:turperiodictwo}
    \mathcal{V}_{\rm \scriptscriptstyle{TUR}}(\omega)=2\left(1-\frac{\omega}{P\ped{out}(\omega)} \frac{\partial{P\ped{out}(\omega)}}{\partial{\omega}}\right)^2,
    \end{equation}
  
   is the dynamic TUR bound.~Eq.~\eqref{eq:tureff} and~\eqref{eq:turperiodic} provide expressions for the bound in terms of experimentally accessible quantities. 
   
   The nonequilibrium dynamics of the expectation values of power operators in Eq.~\eqref{eq:meanpowers}, and the two-time correlation functions as in Eq.~\eqref{eq:averagecorr} for the work-to-work converter in Eq.~\eqref{eq:TotalHamchap3} can be simulated numerically by employing the SIL method (see  App.~\ref{app:SILapp}), for every value of the fields amplitudes.~We can thus compute all the quantities involved in Eq.~\eqref{eq:turperiodic} in order to investigate the validity of TUR.
   In the following section, we focus on the characterization of the converter performance in the linear response regime,~\ie~adopting the framework of linear irreversible thermodynamics \cite{BENENTI20171,Proesman:powereff}.~This formalism allows us to easily find optimal operating regimes for the work-to-work converter.~We postpone the analysis of the converter performance in the nonlinear regime to a subsequent work.
  
   \section{Linear Response regime}\label{sec:linearregime}
	
    The Hamiltonian in Eq.~\eqref{eq:TotalHamchap3} can be rewritten by grouping all the operators which explicitly depend on time $t$ in the following way 
	\begin{equation}\label{eq:perturb}
	\begin{gathered}
	H(t)=H\ped{0} + H\ped{ext}(t)\mbox{,  }\\
	H\ped{ext}(t)=-\sum_{i=1,2} \frac{1}{2}\epsilon_i (t)\sigma\ped{z},
	\end{gathered}
	\end{equation}
	where the operator $H\ped{0}$ is the time-independent contribution to the total Hamiltonian.~In the linear response approach, the term $H\ped{ext}(t)$ acts as a perturbation to the Hamiltonian $H\ped{0}$.~The fluctuation of the expectation value of a generic observable $O(t)$ with respect to the unperturbed case can be written at first order in the perturbation $H\ped{ext}(t)$ as follows  
	\begin{equation}\label{eq:linresp}
	\ev{O(t)} -\ev{O(t)}\ped{0}= i\int\limits_{t_0}^{t}\mathrm{d}t^{\prime}\ev{\comm{H\ped{ext}(t^{\prime})}{O(t)}}\ped{0},
	\end{equation}
	where
	$\ev{O(t)}=\tr\qty[O(t)\rho(t_{0})]$, the subscript $0$ indicates that the expectation values are computed with respect to the unperturbed, time-independent Hamiltonian $H\ped{0}$, and $\rho(t_{0})$ is the state of the system at $t=t\ped{0}$.~Here we want to compute the response of the power operators in Eq.~\eqref{eq:powerst}: inserting the formal expressions of the operators in Eq.~\eqref{eq:linresp}, we obtain
	\begin{equation}\label{eq:finalresp}
	\ev*{P_j(t)} -\ev*{P_j (t)}\ped{0}=\frac{1}{4}\sum_{i=1,2}\int\limits_{t_0}^{+\infty}\mathrm{d}t^{\prime}\dot{ \epsilon}_j (t)\epsilon\ped{i}(t^{\prime})\chi(t,t^{\prime}),
	\end{equation}
	where we recasted Eq.~\eqref{eq:linresp} in terms of the susceptibility $\chi(t,t^{\prime})$
	\begin{equation}
	\chi(t,t^{\prime})=-i\Theta(t-t^{\prime})\tr\qty[\comm{\sigma\ped{z}(t)}{\sigma\ped{z}(t^{\prime})}\rho(t_{0})]\ped{0}.
	\end{equation}
	Here $j=1,2$ and the $\sigma\ped{z}$ operators are computed in the Heisenberg representation of the unperturbed Hamiltonian $H\ped{0}$.~At long times $t$, the correlation function on the right-hand side of Eq.~\eqref{eq:finalresp} becomes a function of the time differences $t-t^{\prime}=\tau$ and the expectation value $\ev*{P_j (t)}\ped{0}$ vanishes.~After straightforward manipulations, we can take the average of the power expectation values over a period $\mathcal{T}$, in order to compute the nonequilibrium stationary mean powers as in Eq.~\eqref{eq:meanpowers} in the limit of linear response.~Introducing the correlation function $C(\tau)=\tr\qty[\sigma\ped{z}(\tau)\sigma\ped{z}(0)\rho(\overline{t})]\ped{0}$, where $\overline{t}$ is a sufficiently long time, we find for the mean powers 
	\begin{equation}\label{eq:finalresp2}
	\overline{P}_j =\frac{1}{2}\sum_{i=1,2}\frac{1}{\mathcal{T}}\int\limits_{0}^{\mathcal{T}}\mathrm{d}t\dot{\epsilon}_{j}(t)\int\limits_{0}^{+\infty}\mathrm{d}\tau\epsilon_{i}(t-\tau)\Im\{C(\tau)\}.
	\end{equation}
	From Eq.~\eqref{eq:finalresp2}, we can easily derive the general form of the Onsager matrix, which links the mean powers $\overline{P}_j$ to the field amplitudes as follows
	\begin{equation}\label{eq:LinOnsager}
	\overline{P}_i (\omega)=\sum_{j=1,2}\mathcal{L}_{ij}(\omega)\epsilon_i\epsilon_j.
	\end{equation}
	In this limit, the elements of the Onsager matrix can be expressed in terms of the Fourier transform of the correlation function $\Im\{C(\tau)\}$,~\ie~we rewrite Eq.~\eqref{eq:finalresp2} in terms of $\mathcal{F}(\omega)=\int_0^{\infty}\exp (i\omega \tau) \Im\{C(\tau)\}\mathrm{d}\tau$.~In what follows, we choose $n=1$,~\ie~the fields oscillate with the same frequency.~Thus, inserting the field expressions $\epsilon_1(t),\epsilon_2(t)$ into Eq.~\eqref{eq:finalresp2}, for each value of the driving frequency $\omega$ and the phase difference $\varphi$, the Onsager matrix reads 
	\begin{equation}\label{eq:Onsagercoeff}
	\begin{gathered}
	\mathcal{L}_{11}(\omega)=\mathcal{L}_{22}(\omega)=-\frac{\omega}{4}\Im \{\mathcal{F}(\omega)\}, \\
	\mathcal{L}_{12}(\omega)=\frac{\omega}{4}(\cos\varphi \Re\{\mathcal{F}(\omega)\} -\sin\varphi \Im\{\mathcal{F}(\omega)\}), \\
	\mathcal{L}_{21}(\omega)=-\frac{\omega}{4}(\cos\varphi \Re\{\mathcal{F}(\omega)\} +\sin\varphi \Im\{\mathcal{F}(\omega)\} ).
	\end{gathered}
	\end{equation}
	Notice that when the system operates in TR symmetric regime,~\ie~$\varphi=\pi/2$, the Onsager matrix is symmetric, and all its element are equal to one another.~The knowledge of Onsager functions allows us to access the mean fluctuations of the output power.~Following a similar treatment as in Eq.~\eqref{eq:finalresp}, starting from Eq.~\eqref{eq:averagecorr} we can write
	\begin{align}\label{eq:meanfluct}
	\overline{D}_i=\frac{\epsilon^2_i \omega^2}{4}\int\limits_0^{+\infty}\mathrm{d}\tau \cos (\omega \tau)\Re\{C(\tau)\}.
	\end{align}
	The last expression can be easily written in terms of the diagonal elements of the Onsager matrix. Considering the properties of $C(\tau)$, at long times $\tau$ we have 
	\begin{align}\label{eq:meanfluct2}
	\overline{D}_i=\epsilon^2_i \omega \coth(\frac{\beta \omega}{2}) \mathcal{L}_{ii}(\omega). 
	\end{align}
	From Eq.~\eqref{eq:finalresp2} and \eqref{eq:meanfluct2}, it follows that the computation of the correlation function $C(\tau)$ allows us to characterize the nonequilibrium dynamics of the converter in the linear response regime.~We stress that, although the function $C(\tau)$ is computed from the interacting spin-boson Hamiltonian in the absence of external fields, analytical solutions for $C(\tau)$ are limited to special values of the dissipation strengths \cite{SassettiWeiss:SpinBoson} (see Sec.~\ref{subsec:Toulouse} and App.~\ref{app:weakcoup}); as a consequence, even in the linear response regime, a fully numerical approach is required.       

	Once the Onsager matrix is known, optimal working conditions can be easily found with straightforward algebra \cite{BENENTI20171}.~The line of maximum efficiency (ME) in the parameter space $(\epsilon_1,\epsilon_2)$, for fixed driving frequency and phase difference $(\omega,\varphi)$ reads
	\begin{equation}\label{eq:maxeffline}
    \epsilon_{1 \rm{\scriptscriptstyle{ME}}}= \epsilon_2\frac{\mathcal{L}_{22}(\omega)}{\mathcal{L}_{21}(\omega)}\qty(\frac{1}{\sqrt{1+\mathcal{Y}}} -1).
	\end{equation}
	
	Here $\mathcal{Y}=\mathcal{L}_{12}\mathcal{L}_{21}/\det\mathcal{L}$ and       $\det\mathcal{L}$ 
	denotes the determinant of the Onsager matrix.~The converter figure of merit, the output power and the relative power uncertainty can be computed at ME starting from Eq.~\eqref{eq:eff}, \eqref{eq:LinOnsager}, \eqref{eq:meanfluct2}.~By introducing the asymmetry factor $\mathcal{X}=\mathcal{L}_{12}/\mathcal{L}_{21}$, they read respectively
    \begin{equation}\label{eq:perfme}
    \begin{gathered}
     \eta_{\rm{\scriptscriptstyle{ME}}}=\mathcal{X}\frac{\sqrt{1+\mathcal{Y}}-1}{\sqrt{1+\mathcal{Y}}+1},  \\
      P_{\rm{out,\scriptscriptstyle{ME}}}=\overline{P}_{1, \rm{\scriptscriptstyle{ME}}}=-\epsilon^2_2 \eta_{\rm{\scriptscriptstyle{ME}}}\frac{\mathcal{L}_{22}(\omega)}{\sqrt{1+\mathcal{Y}}},\\
      \Sigma^2_{\rm{out,\scriptscriptstyle{ME}}}=\omega \frac{\mathcal{L}_{11}(\omega)}{\mathcal{L}^2_{12}(\omega)}\qty(1+\sqrt{1+\mathcal{Y}})^2\coth(\frac{\beta\omega}{2}).
    \end{gathered}
    \end{equation}	

   \section{Converter performance and TUR violation}\label{sec:violations}
   
   Below, we simulate numerically the dynamics of the converter. We employ the numerical SIL approach, which allows us to compute the unitary dynamics of the driven TLS + bath density operator $\rho(t)$, after a controlled truncation of the bath Hilbert space, without recurring to further approximations (see  App.~\ref{app:SILapp}). In the linear response regime, we first identify regions in the parameter space where the system operates as a work-to-work converter; then, by choosing the operating point at ME, we characterize the performance of our converter,~\ie~we compute output power, efficiency and fluctuations, for different values of the coupling strength $\alpha$, in the low temperature regime where non-Markovian effects and quantum coherence are expected. Eventually, we compute both sides of Eq.~\eqref{eq:turperiodic} and indentify several frequency intervals where the Markovian TUR cannot hold.~We also study the very special case of $\alpha=1/2$, where the converter performance can be computed analytically.

   \subsection{Converter dynamics, energy exchange}\label{subsec:convdyn}
   
   We set the density matrix of system and the bath at initial time $t\ped{0}$ in a factorized state as follows 
   \begin{equation}\label{eq:startcond1}
   \rho(t\ped{0})=\rho\ped{S}(t\ped{0})\otimes \frac{e^{-\beta H\ped{B}}}{Z\ped{B}},
   \end{equation}       
   where $\rho\ped{S}(t\ped{0})=\ket{+}\bra{+}$.~Thus, we simulate the nonequilibrium dynamics of the expectation values of one time and two-time operators of interest, for different values of the model parameters, fixing the maximum number of excitations to $N\ped{ph}=3$ and the number of bath modes $M\ped{mod}=220$ (see App.~\ref{app:SILapp}).~We also fix the phase difference $\varphi=0$,~\ie~the converter operates in a TR asymmetric configuration.~We 
   \begin{figure}[h]
    	\centering
    	\includegraphics[width=0.99\linewidth]{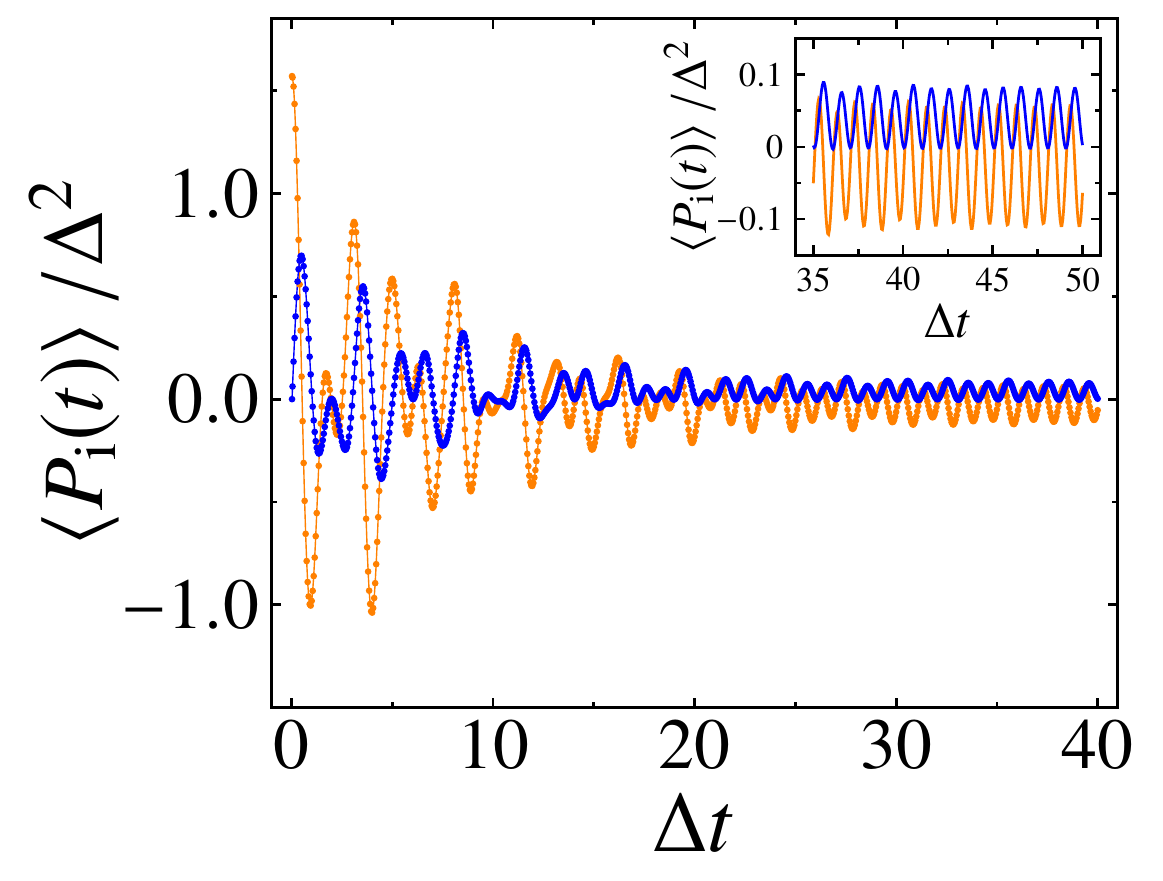}
    	\caption{Expectation value of power operators $\ev{P_{i}(t)}/\Delta^2\mbox{,  }i=1\text{ (orange), }i=2\text{ (blue)}$ as function of time $t$, with $\epsilon_1=-\Delta\mbox{, }\epsilon_2=0.50\Delta\mbox{, }\varphi=0\mbox{, }\alpha=0.10\mbox{, }T=0\mbox{, }\omega\ped{c}=10\Delta\mbox{, }N\ped{ph}=3\mbox{, }M\ped{mod}=220$.~Inset: nonequilibrium stationary state at long time shows mean powers oscillating with period $\mathcal{T}$, and the time averages are of opposite signs, signaling work-to-work conversion.}
    	\label{fig:powersenergy1} 
   \end{figure} 
    \begin{figure}[h]
   	\centering
   	\includegraphics[width=0.95\linewidth]{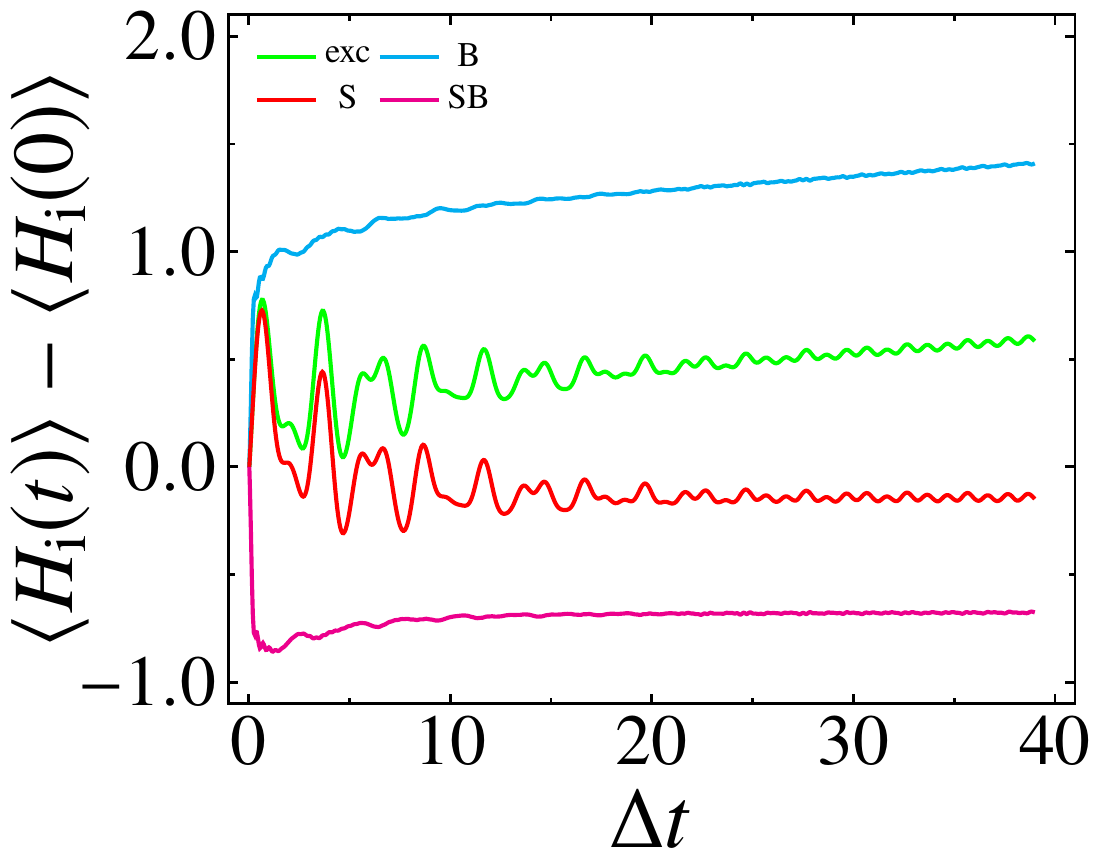}
   	\caption{Time evolution of the variation in the expectation values of energy operators written in Eq.~\eqref{eq:TotalHamchap3},~\ie~$\ev{H\ped{i}(t)}-\ev{H\ped{i}(0)}$ where $i=\text{S},\text{B},\text{SB}$ and $\text{exc}$ denote the TLS, the bath, the interaction energies and the total energy pumped into the system respectively.~The model parameters are fixed as follows: $\epsilon_1=-\Delta\mbox{, }\epsilon_2=0.50\Delta\mbox{, }\varphi=0\mbox{, }\alpha=0.10\mbox{, }T=0\mbox{, }\omega\ped{c}=10\Delta\mbox{, }N\ped{ph}=3\mbox{, }M\ped{mod}=220$.~It can be noticed that the total power is entirely drained by the bath.}
   	\label{fig:powersenergy2} 
   \end{figure}  
   start by describing the main features of the nonequilibrium dynamics of the powers expectation values in Eq.~\eqref{eq:powerst} and the energy exchange of the converter.~In Fig.~\ref{fig:powersenergy1}, the dynamics of the expectation values of powers in the two channels $\ev{P_i (t)}$ is plotted for fixed values of the fields parameters, dissipation strength and temperature.~It is shown that, while the powers exhibit a transient behavior marked by fast oscillations in time of decreasing amplitude, for $t\geq 30\Delta\ped{eff}^{-1}$ a stationary regime sets in: here the power expectation values undergo periodic oscillations in time, with period equal to $\mathcal{T}$, the mean values over a period being different from zero and of opposite signs.~A part of the work spent per unit time in a channel is thus given back in the other,~\ie~the system operates as a work-to-work converter with efficiency $\eta$.  
  
   The analysis of energy exchange among the TLS, the bath and the interactions channel can be performed by plotting the expectation values of the different contributions in Eq.~\eqref{eq:TotalHamchap3} as function of time, as reported in Fig.~\ref{fig:powersenergy2}.~Notice that also the different energy contributions experience a transient behavior and, for sufficiently long times a nonequilibrium stationary state is reached, where they oscillate in time with period $\mathcal{T}$.~However, the analysis confirms that the mean total power $P\ped{exc}=P\ped{in}+P\ped{out}$ is entirely drained by the bath while the TLS and the interaction energies oscillate around constant values, and the mean power drained by their channels vanishes.~We stress that Fig.~\ref{fig:powersenergy1} and~\ref{fig:powersenergy2} report the numerically exact expectation values of the operators of interest.~It is a clear advantage of our numerical approach (see App.~\ref{app:SILapp}), which can be directly employed in both the linear and nonlinear regime.~From Fig.~\ref{fig:powersenergy2}, it follows that the energy exchange has a clear interpretation only in the nonequilibrium steady state, while during the transient time a nontrivial energy exchange mechanism among the three different channels shows up.
   
   The converter dynamics can also be characterized by means of a measure of non-Markovianity for open quantum systems \cite{Breuer:NonMarkov}.~This measure can be defined for the reduced state $\rho\ped{S}(t)$ of the open quantum system, by employing the notion of distinguishability of quantum states.~The trace distance between couples of reduced states of the system is thus introduced as
   \begin{equation}\label{eq:NonMarkov} 
   D(\rho\ped{S}^{\scriptsize{(1)}}(t),\rho\ped{S}^{\scriptsize{(2)}}(t))=\frac{1}{2}\Tr\sqrt{(\Delta \rho)^{\dagger}(\Delta \rho)}, 
   \end{equation}
   where $\Delta\rho=\rho\ped{S}^{\scriptsize{(1)}}(t)-\rho\ped{S}^{\scriptsize{(2)}}(t)$.~The existence of at least a couple of initial states of the reduced system $\rho\ped{S}^{\scriptsize{(1)}}(0),\rho\ped{S}^{\scriptsize{(2)}}(0)$ such that $\mathrm{d}D(\rho_1(0),\rho_2(0),t)/\mathrm{d}t>0$ in a given interval of time implies that the dynamics of the reduced state is non-Markovian,~\ie~information does not flow monothonically through the reservoir, but information backflows during the dynamics can be observed.~The actual degree of non-Markovianity can be defined in different ways starting from Eq.~\eqref{eq:NonMarkov} \cite{Breuer:NonMarkov,Poggi:NonMarkov}.~In    
   \begin{figure}[h]
   	\centering
   	\includegraphics[width=0.95\linewidth]{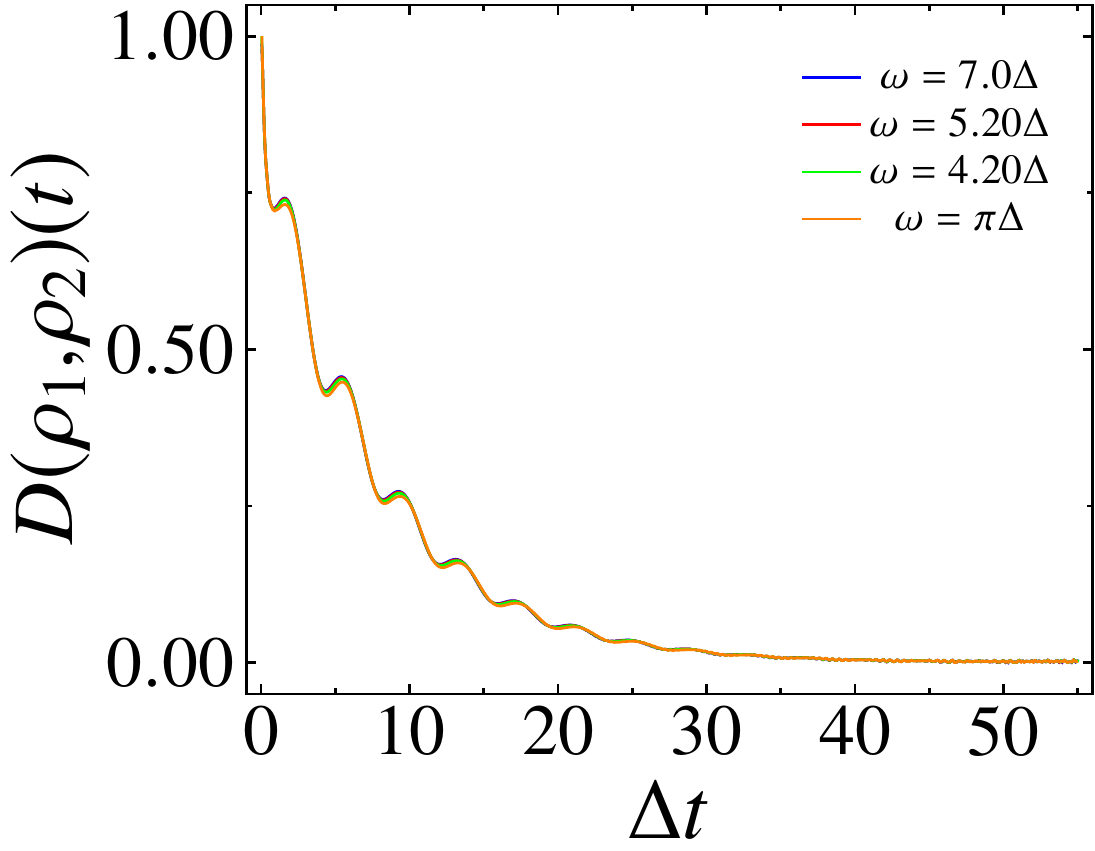}
   	\caption{Time evolution of the trace distance as in Eq.~\eqref{eq:NonMarkov}, having fixed the initial conditions $\rho^{(1)}_0=\ket{y,+}\bra{y,+}\mbox{, }\rho^{(2)}_0=\ket{y,-}\bra{y,-}$, for different driving frequencies $\omega$ in the range $[\pi\Delta,7.0\Delta]$.~The model parameters have been fixed as follows: $\epsilon_1=\epsilon_2=0.50\Delta\mbox{, }\varphi=0\mbox{, }\omega\ped{c}=10\Delta\mbox{, }\alpha=0.10\mbox{, }T=0\mbox{, }N\ped{ph}=3\mbox{ and }M\ped{mod}=220$.}
   	\label{fig:nonmarkov}
   \end{figure}
   Fig.~\ref{fig:nonmarkov}, we plot the behavior of the non-Markovianity witness reported in Eq.~\eqref{eq:NonMarkov} as function of time, for fixed values of model parameters and different initial states.~It is evident that during the transient time a noticeable non-Markovian behavior can be observed, as follows from the nonmonotonicity of the trace distance.~However, as the nonequilibrium stationary state is reached, the reduced system states become progressively less distinguishable.~This result points toward a limited influence of non-Markovian effects, at least in the steady state operating regime of our converter.
   
   \subsection{Performance in the linear response regime}\label{subsec:conveff}
  
   Limiting the analysis to the linear response regime, from the expression of the Onsager matrix elements reported in Eq.~\eqref{eq:Onsagercoeff} for fixed phase difference $\varphi$, the occurrence of several frequency regions where work-to-work conversion is present follows. 
   
\begin{figure}[h]
   	\centering
   	\includegraphics[width=0.99\linewidth]{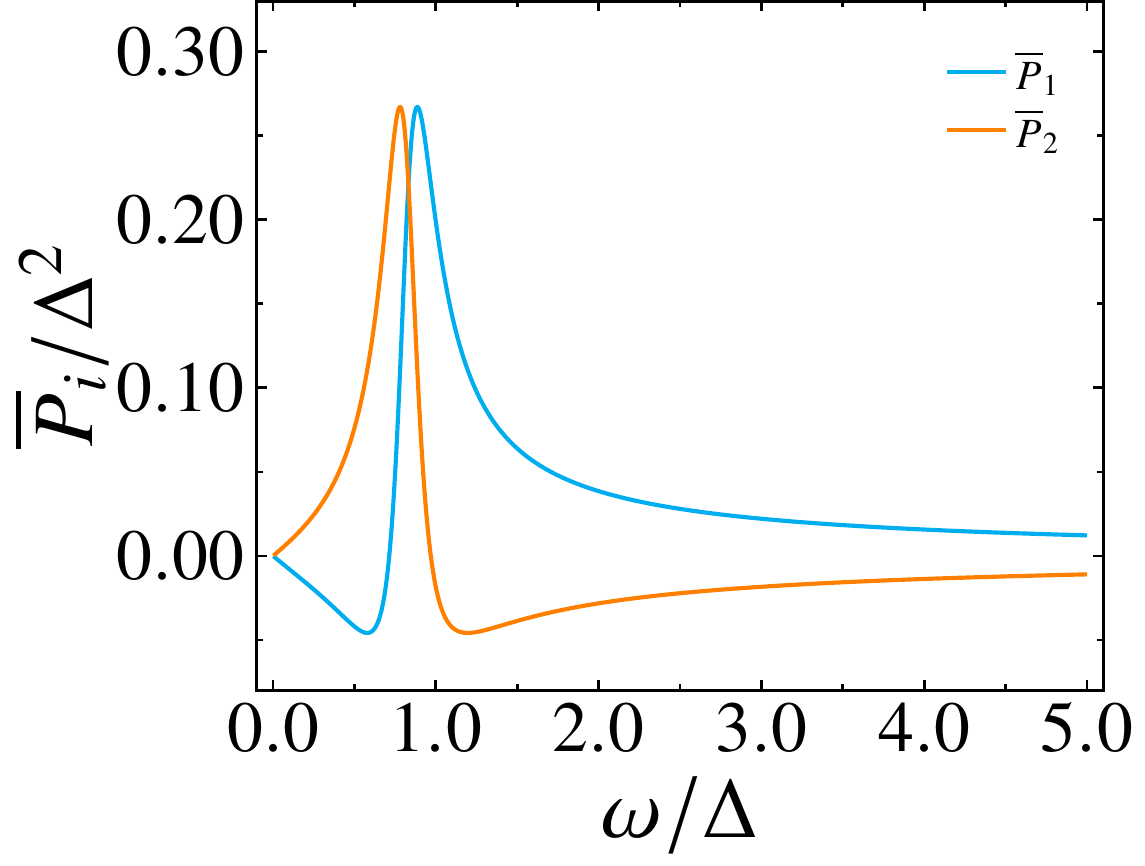}
   	\caption{Average nonequilibrium powers $\overline{P}_1,\overline{P}_2$ computed in the linear response regime as function of the driving frequency $\omega$.~The model parameters have been fixed as follows: $\epsilon_1=\epsilon_2=0.50\Delta\mbox{, }\varphi=0\mbox{, }\alpha=0.10\mbox{, }T=0.10\Delta\mbox{, }\omega\ped{c}=10\Delta$.~It is evident that a distinct region of driving frequencies exists so that the system cannot operate as work-to-work converter.}
   	\label{fig:Onsag}
   \end{figure}

   With $\varphi=0$, the Onsager matrix in Eq.~\eqref{eq:Onsagercoeff} is antisymmetric.~In Fig.~\ref{fig:Onsag}, we plot the mean powers related to channels $1,2$ against the driving frequency $\omega$, as derived from Eq.~\eqref{eq:LinOnsager}, for fixed dissipation strength and temperature $T$: it can be noticed that they both hold positive for frequencies of the order of the TLS oscillation frequency $\Delta\ped{eff}$ (see App.~\ref{app:weakcoup}),~\ie~when the driving frequency is near resonance.~Here the system absorb the power injected along the two channels,~\ie~the heat flux to the bath is maximum and no work conversion occurs. 
   On the other hand, when the system is driven with frequency sufficiently far from resonance, the conversion takes place with finite efficiency.~These results clearly depend on the phase difference $\varphi$ between the two drives,~\ie~the parameter controlling the TR asymmetry of the system.  

  	\begin{figure}[h]
  	\centering
  	\includegraphics[width=0.9999\linewidth]{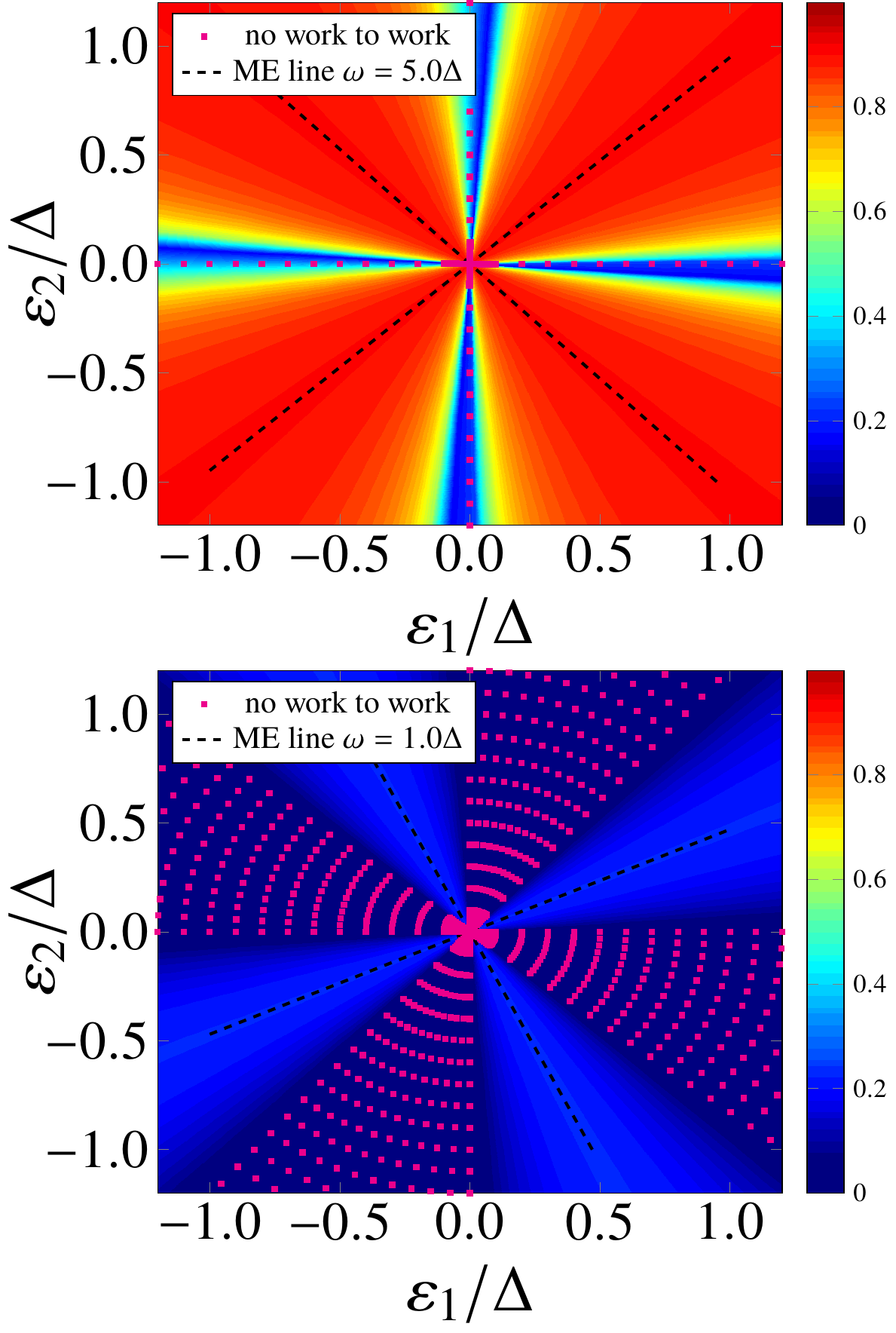}
  	\caption{Density plot of the efficiency as a function of the driving fields amplitudes $(\epsilon_1,\epsilon_2)$, plotted for fixed $\alpha=0.10$, $T=0.10\Delta$, $\varphi=0$ and $\omega=\{\Delta,5\Delta\}$ respectively.~Magenta dots mark the parameter regions where no work-to-work conversion occurs, while the $\text{ME}$ line in Eq.~\eqref{eq:maxeffline} is plotted (dashed line).}
  	\label{fig:colormaps} 
  \end{figure}
  
  The conversion efficiency $\eta$ can be computed from Eq.~\eqref{eq:eff} as a function of the field amplitudes $(\epsilon_1,\epsilon_2)$, for fixed dissipation strength and temperature.~In Fig.~\ref{fig:colormaps} we plot the efficiency for two different values of the driving frequency $\omega=\{\Delta,5\Delta\}$.~It is evident that, tuning the driving frequency near the resonance, all but limited regions of the parameters space $(\epsilon_1,\epsilon_2)$ exhibit no work-to-work conversion, and in the regions where conversion occurs a very low efficiency is achieved.~On the other hand, tuning the frequency out of resonance a very different scenario can be observed, where the efficiency reaches near-to-one values along the ME lines in Eq.~\eqref{eq:maxeffline} (dashed lines in Fig.~\ref{fig:colormaps}), while small but finite input and output powers can be achieved, as shown in Fig.~\ref{fig:Onsag}.
   
  As a consequence, choosing the converter operating point along the ME line allows us to optimize its performance.~However, the analysis of the fluctuations of the output powers in this regime is also required, as with any microscopic heat engine.~In addition, the mean fluctuations enter directly in the definition of TUR in Eq.~\eqref{eq:turperiodic}.~Moreover, increasing the strength of dissipation at fixed temperature the converter performance degrades, and output power, efficiency and fluctuations as functions of the driving frequency are severely altered. 
 
\begin{figure}[h!]
   	\centering 
   	\includegraphics[width=8.6 cm]{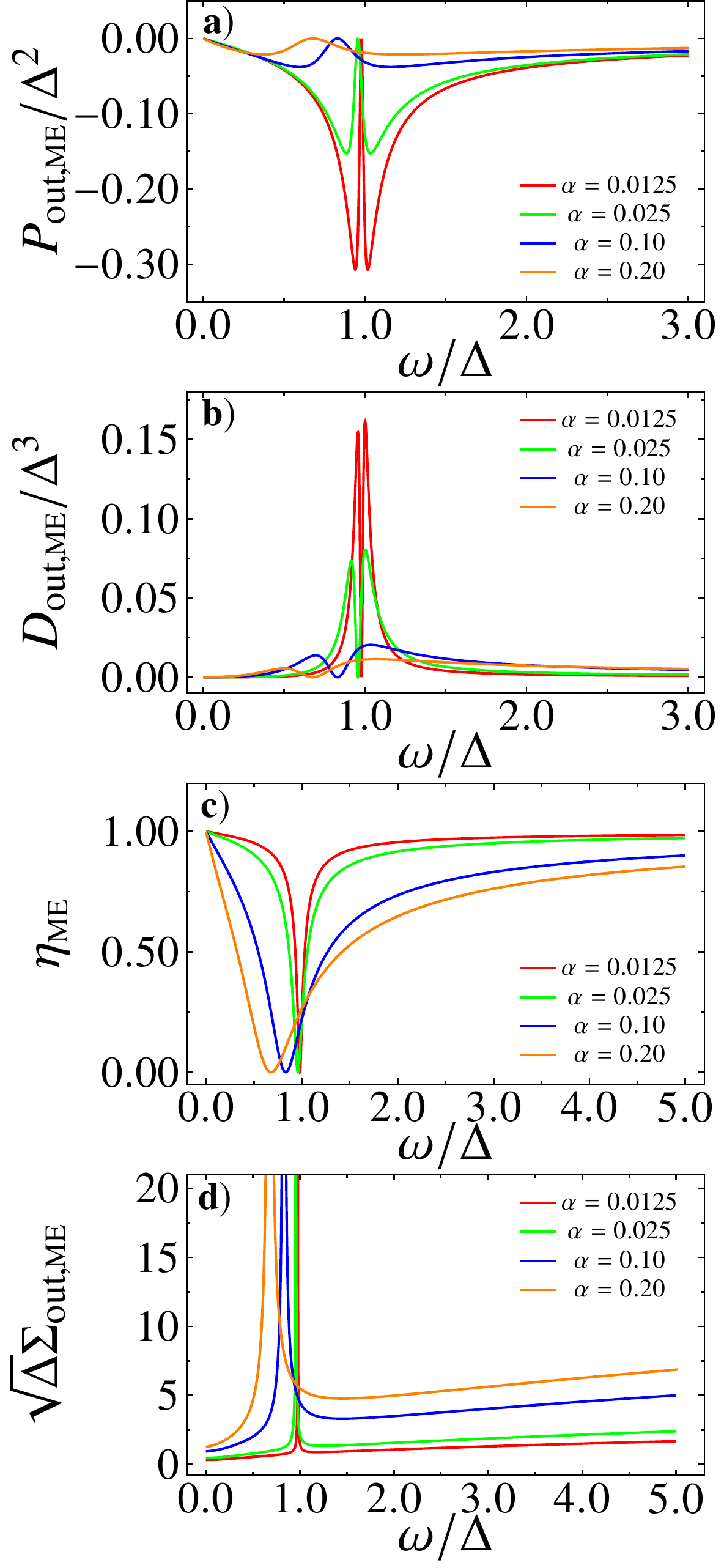}
   	\caption{Performance of the work-to-work converter at ME as a function of the driving frequency $\omega$, for different values of the dissipation strength $\alpha$ and fixed temperature $T=0.10\Delta$, frequency cutoff $\omega\ped{c}=10\Delta$ and $\epsilon_2=0.50\Delta$.~In Panel $\mathbf{a}$, the output power $P\ped{out}$.~In Panel $\mathbf{b}$, the output power fluctuations; in Panel $\mathbf{c}$, the efficiency curve computed from Eq.~\eqref{eq:eff}; in Panel $\mathbf{d}$, the relative power uncertainty $\Sigma_{\rm{out,\scriptscriptstyle {ME}}}=\sqrt{D_{\rm{out,\scriptscriptstyle {ME}}}/P^2_{\rm{out,\scriptscriptstyle {ME}}}}$.~The phase difference has been fixed to $\varphi=0$, so that the chosen point along ME line in Eq.~\eqref{eq:maxeffline} remains fixed for each value of $\omega$.}
   	\label{fig:Performance}
    \end{figure}

  We thus set the operating point of the converter on the ME line in Eq.~\eqref{eq:maxeffline} and investigate the effects of dissipation on the converter performance, for any frequency value $\omega$, at fixed temperature.~In Fig.~\ref{fig:Performance}, we report the plots of the output power $P_{\rm{out,\scriptscriptstyle {ME}}}$
  (panel $\mathbf{a}$), mean fluctuations $D_{\rm{out,\scriptscriptstyle {ME}}}$ (panel $\mathbf{b}$), efficiency $\eta_{\rm{\scriptscriptstyle {ME}}}$ (panel $\mathbf{c}$) and relative power uncertainty $\Sigma_{\rm{out,\scriptscriptstyle {ME}}}$ (panel $\mathbf{d}$) at ME as function of the driving frequency, for fixed temperature $T=0.10\Delta$ and different values of the coupling strength $\alpha$ taken in the range $\numrange[range-phrase = \text{~to~},exponent-product=\cdot]{1.25e-2}{2.0e-1}$.   
   The powers in the two channels at ME are quite different from those in Fig.~\ref{fig:Onsag}, as they don't change sign crossing the resonance.~The plot of the output power as function of driving frequency $\omega$ shows a characteristic double-peak feature, marked by a narrow region where it drops to zero.~By increasing the driving frequency $\omega$, the output as well as the input power smoothly decreases.~For increasing dissipation strengths, the double-peaked structure tends to smooth down and the resonance frequency, due to progressive shrinking of the TLS gap, moves toward lower frequencies.~The input power changes similarly as a function of dissipation strength $\alpha$.~As a direct consequence of this behavior, it can be noticed that the maximum efficiency curve $\eta_{\rm{\scriptscriptstyle {ME}}}$ decreases with increasing $\alpha$ in the whole frequency region,~\ie~the entropy production grows as a function of the dissipation strength.~However, even in the presence of moderately strong dissipation,~\ie~$\alpha<0.2$ the efficiency retains much of its shape.~For increasing frequencies $\omega$ its value remains above $0.50$, while in the low frequency region it tends to $1$.
   
   Mean fluctuations of the output power $D_{\rm{out,\scriptscriptstyle {ME}}}/\Delta^3$ have also been shown as function of the frequency $\omega$.~Notice that, in the resonance region their shape is quite similar to the output power, though the double-peaked structure shows evident asymmetry.~Moreover, fluctuations behave very differently at high driving frequencies with respect to lower ones.~Above resonance, a marked difference with respect to the output power can be observed: for $\omega>\Delta\ped{eff}$ and for increasing coupling strength $\alpha$, a distinct growth in the mean fluctuations can be noticed.~This feature, along with the increase in the entropy production and the renormalization of the tunnelling element, is a characteristic effect of the quantum dissipative environment on the working medium.~By inspecting the mean relative uncertainty, it is shown that a divergence occurs in the vicinity of resonance region, mainly due to the rapid drop of output power with respect to mean fluctuations.~Further, for increasing driving frequencies the relative uncertainty falls to a minimum and then start to grow slowly,~\ie~at high frequency the converter progressively loses precision.~The loss of precision in the conversion process increases mainly due to the environment effect.           
   
   In the weak coupling regime, it has been shown (see Fig.~\ref{fig:Performance}, panel $\mathbf{c}$) that $\eta\to 1$ in the low and high driving frequency regions, and correspondingly the power output decreases.~However, it is interesting to show the detailed behavior of $\abs{P_{\rm{out,\scriptscriptstyle {ME}}}}$ and $D_{\rm{out,\scriptscriptstyle {ME}}}$ as function of $1-\eta_{\rm{\scriptscriptstyle {ME}}}$ in the same limits.~In
    \begin{figure}[h]
   	\centering
   	\includegraphics[width=0.99\linewidth]{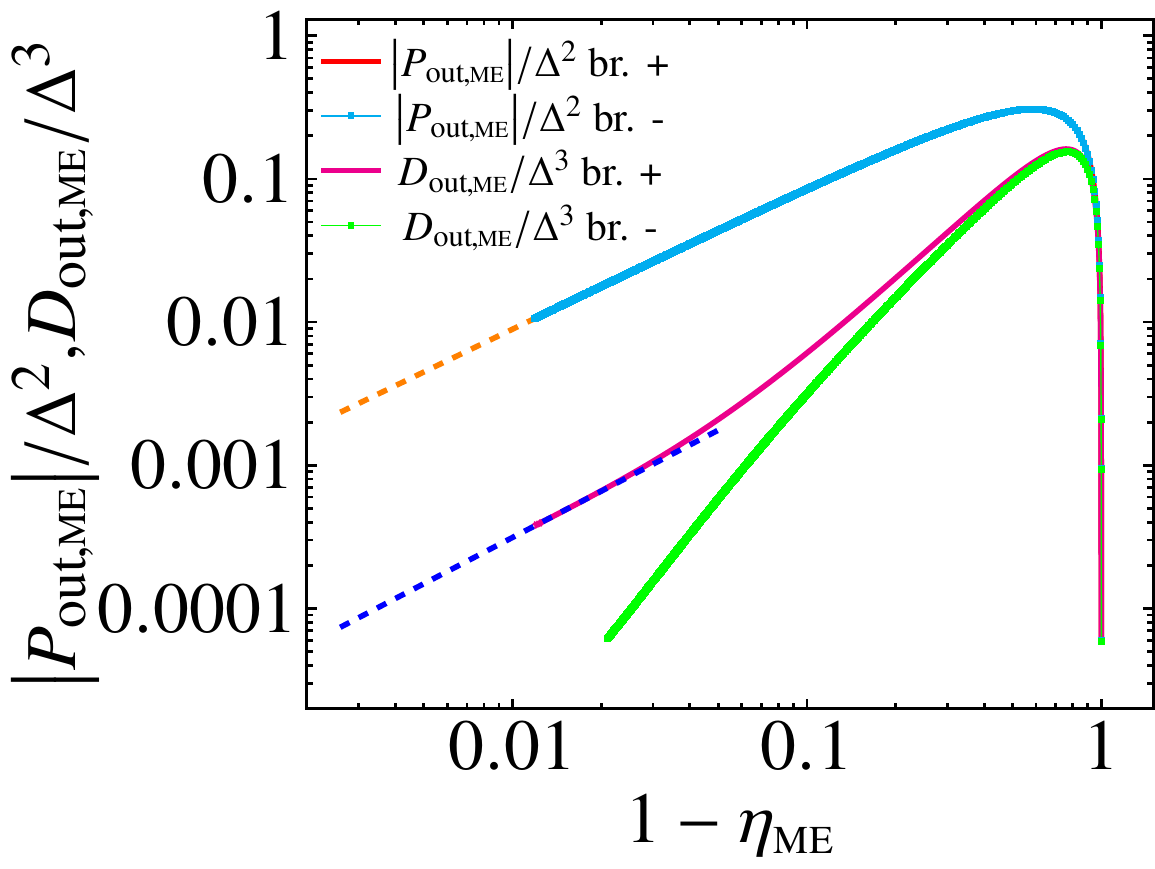}
   	\caption{Mean output power and fluctuations at ME $\abs{P_{\rm{out,\scriptscriptstyle {ME}}}}/\Delta^2\mbox{, }D_{\rm{out,\scriptscriptstyle {ME}}}/\Delta^3$ computed in the linear response regime as function of $1-\eta_{\rm{\scriptscriptstyle {ME}}}$, for fixed parameter values $\epsilon_2=0.50\Delta\mbox{, }\varphi=0\mbox{, }\alpha=0.0125\mbox{, } \beta=10\Delta^{-1}$, in the positive ($+$) and negative ($-$) branch.~For $\eta\to 1$, in the $+$ branch the powers and the mean fluctuations drop linearly with $1-\eta_{\rm{\scriptscriptstyle {ME}}}$, while in the $-$ branch mean fluctuations decay faster. The dashed lines denote the fitted asymptotic behaviors of the output power (orange) and the mean fluctuations at ME in the positive branch (blue).} 
   \label{fig:powerlaw}  
   \end{figure}
   Fig.~\ref{fig:powerlaw}, $\abs{P_{\rm{out,\scriptscriptstyle {ME}}}}/\Delta^2,D_{\rm{out,\scriptscriptstyle {ME}}}/\Delta^3$ are reported as a function of $1-\eta\ped{\scriptsize{ME}}$, for fixed coupling strength $\alpha=0.0125$, in the positive and negative branches,~\ie~ for frequencies above and below resonance respectively.~It can be seen that the output power vanishes linearly with respect to $1-\eta_{\rm{\scriptscriptstyle {ME}}}$, with the same slope in the positive and negative branches.~Actually this result can be derived from the analytic expression of the Onsager matrix in the limit of weak coupling, as reported in App.~\ref{app:weakcoup}.~It can be shown that at fixed temperature $T$, the slopes of $\abs{P_{\rm{out,\scriptscriptstyle {ME}}}}\mbox{, }D_{\rm{out,\scriptscriptstyle {ME}}}$ curves in the positive branch can be captured by two different, monotonically decreasing functions of the coupling strength $\alpha$, as reported in Eq.~\eqref{eq:linslope}.~From Fig.~\ref{fig:powerlaw}, it can also be noticed that, in the negative branch, mean power fluctuations curves vanish more quickly than in the positive branch.
 
   It is interesting to remark that the power-efficiency trade-off shown in Fig.~\ref{fig:powerlaw} is in agreement with the optimal dependence predicted \cite{Shiraishi:untradeoff} for classical heat engines whose interactions with heat baths can be described as Markov processes: $\abs{P_{\rm{out,\scriptscriptstyle {ME}}}}\sim \eta_{\rm \scriptscriptstyle C}-\eta$ when $\eta\to \eta_{\rm \scriptscriptstyle C}$ (for isothermal engines, the maximum efficiency is equal to $1$ rather than to $\eta_{\rm \scriptscriptstyle C}$).~While such behavior has already been observed in other models \cite{luo2018}, it is interesting to observe that we are here considering a quantum model where the degree of coherence of the TLS plays an important role and where non-Markovian effects cannot be a priori neglected. Our results may suggest a broader validity range of the results of \cite{Shiraishi:untradeoff}.

   \subsection{TUR violation}\label{subsec:turvio}
   
   The study of converter performance as a function of the driving frequency and dissipation strength, carried out in the previous section, allows us to investigate the validity of the static TUR in Eq.~\eqref{eq:tur} as well as the dynamic TUR in Eq.~\eqref{eq:turperiodic} in the quantum domain.~We compute the tradeoff parameter at ME $\mathcal{Q}_{\rm{\scriptscriptstyle {ME}}}(\omega)$,~\ie~the left-hand side of Eq.~\eqref{eq:tureff}, which links the conversion efficiency $\eta$, the output power $P\ped{out}$ and the relative uncertainty $\Sigma\ped{out}$, as a function of the driving frequency $\omega$, for different values of the coupling strength $\alpha$ and for fixed temperature $T=0.10\Delta$.~We 
       \begin{figure}[h!]
   	\centering
   	\includegraphics[width=7.75 cm]{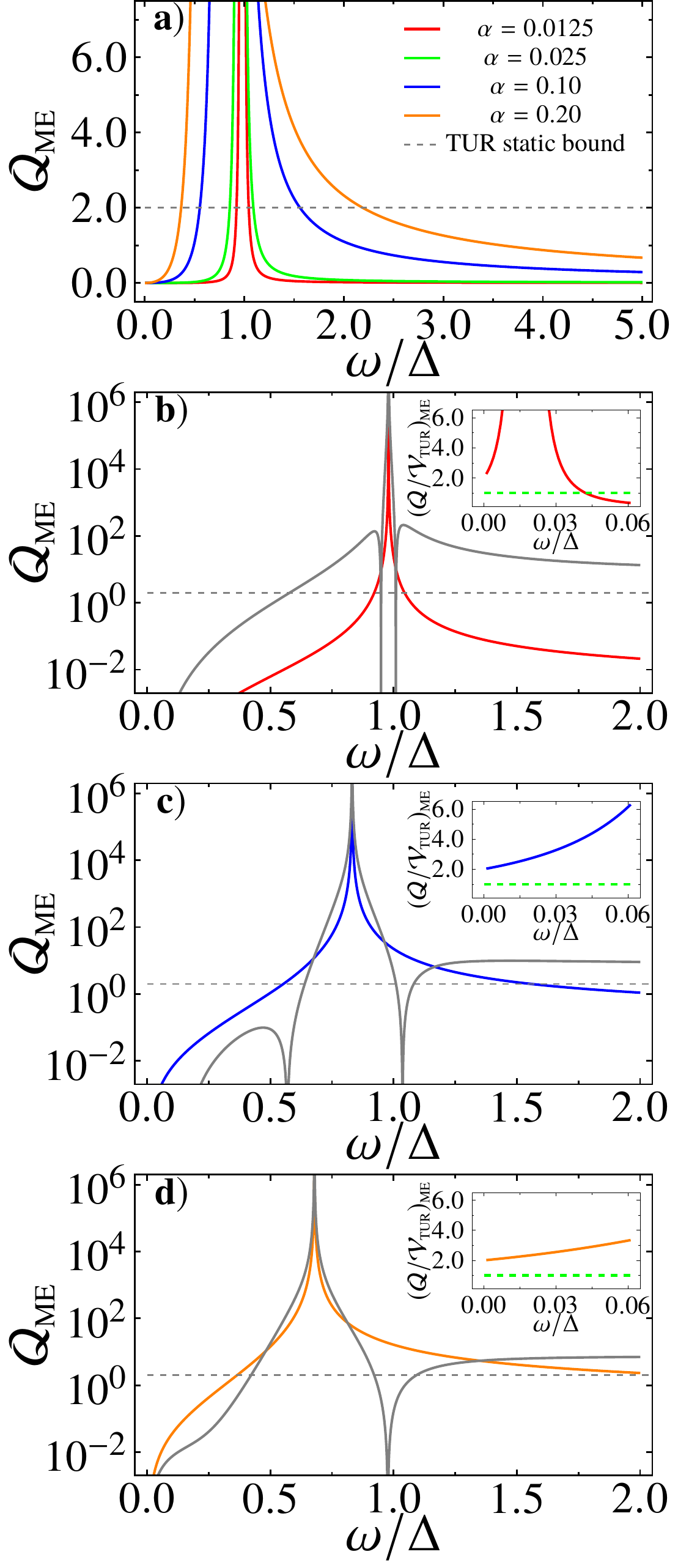}
   	\caption{Tradeoff parameter $\mathcal{Q}_{\rm{\scriptscriptstyle {ME}}}(\omega)$ plotted as a function of the driving frequency $\omega$, for different values of the coupling strength $\alpha$, and compared with the lower TUR bounds in Eq.~\eqref{eq:tur} and~\eqref{eq:turperiodic}.~The temperature is $T=0.10\Delta$, the frequency cutoff $\omega\ped{c}=10\Delta$ and $\epsilon_2=0.50\Delta$ and $\varphi=0$.~In panel $\mathbf{a}$, $\mathcal{Q}_{\rm{\scriptscriptstyle {ME}}}(\omega)$ is plotted against the static TUR bound (dashed gray line), for different values of $\alpha=\{0.0125\mbox{,}0.025\mbox{,}0.10\mbox{,}0.20\}$.~In panels $\mathbf{b}$-$\mathbf{d}$, $\mathcal{Q}_{\rm{\scriptscriptstyle {ME}}}(\omega)$ is plotted against $\mathcal{V}_{\rm{\scriptscriptstyle {TUR,ME}}}(\omega)$ (solid gray line), for $\alpha=0.0125\text{ (solid red line), }\alpha=0.10\text{ (solid blue line), } \alpha=0.20\text{ (solid orange line)}$ respectively.~In the insets of panels $\mathbf{b}$-$\mathbf{d}$, the ratio $(\mathcal{Q}/\mathcal{V}_{\rm{\scriptscriptstyle {TUR}}})_{\rm{\scriptscriptstyle {ME}}}(\omega)$ is plotted in the low-frequency region for each different value of $\alpha$ (green, dashed line signals  $(\mathcal{Q}/\mathcal{V}_{\rm{\scriptscriptstyle {TUR}}})_{\rm{\scriptscriptstyle {ME}}}=1$.)
   	}
   	\label{fig:dynboundcomp} 
   \end{figure}
   compare $\mathcal{Q}_{\rm{\scriptscriptstyle {ME}}}(\omega)$ with the lower TUR bounds,~\ie~with the right-hand side of Eq.~\eqref{eq:tureff} and Eq.~\eqref{eq:turperiodic}, holding in static and periodically-driven external fields case respectively.~In Fig.~\ref{fig:dynboundcomp}, panel $\mathbf{a}$, we show the tradeoff parameter $\mathcal{Q}_{\rm{\scriptscriptstyle {ME}}}(\omega)$ as a function of the driving frequency $\omega$, computed for different values of dissipation strength.~In the resonance region, a divergence of $\mathcal{Q}_{\rm{\scriptscriptstyle {ME}}}(\omega)$ can be observed, which can be traced back to the peculiar behavior of the power uncertainty reported in Fig.~\ref{fig:Performance}, while the entropy production is finite.~In this frequency region, where the system behaves as a trivial dissipator with no work-to-work conversion, the converter performance obeys static TUR.~However, moving away from the resonance region, the tradeoff parameter falls well below the static bound reported in Eq.~\eqref{eq:tur}, thus violating the static TUR in a wide range of out-of-resonance driving frequencies.   

    However, the observed violation might be due to the fact that Eq.~\eqref{eq:tur} doesn't hold for periodically-driven systems,~\ie~it has been derived for Markovian systems subject to static fields \cite{Pietzonka:Tradeoff}.~In Fig.~\ref{fig:dynboundcomp}, panels $\mathbf{b}$-$\mathbf{d}$, we thus compare $\mathcal{Q}_{\rm{\scriptscriptstyle {ME}}}(\omega)$ with the generalized bound  
    $\mathcal{V}_{\rm{\scriptscriptstyle {TUR,ME}}}(\omega)$ in Eq.~\eqref{eq:turperiodic} that has been explicitly derived in the context of periodically-driven Markovian systems.~It is evident that $\mathcal{V}_{\rm{\scriptscriptstyle {TUR,ME}}}(\omega)$ diverges in the resonance region, as $\mathcal{Q}_{\rm{\scriptscriptstyle {ME}}}(\omega)$ does: it is due to the vanishing of the output power reported in Fig.~\ref{fig:Performance}, panel  $\mathbf{a}$.~Moreover, it can be noticed that for $\omega \to 0$, $\mathcal{V}_{\rm{\scriptscriptstyle {TUR,ME}}}(\omega)$ vanishes,~\ie~it does not reduce to the static TUR bound: it follows from the discrete nature of the energy levels of the working medium and from the absence of output and input powers in the static limit.
    Further, from Fig.~\ref{fig:dynboundcomp} a clear dependence of $\mathcal{V}_{\rm{\scriptscriptstyle {TUR,ME}}}(\omega)$ on the dissipation strength can be inferred, so that the frequency regions where dynamic TUR violation occurs change.~Different scenarios can thus be described in the frequency regions below and above resonance.
      
    We first consider the lower frequency region.~In the quasi-static limit, dynamic TUR in Eq.~\eqref{eq:turperiodic} is obeyed for all values of the dissipation strength (see the inset of panel $\mathbf{b}$-$\mathbf{d}$).~At fixed angle $\varphi$, the ratio of $\mathcal{Q}_{\rm{\scriptscriptstyle {ME}}}(\omega)/\mathcal{V}_{\rm{\scriptscriptstyle {TUR,ME}}}(\omega)$ tends to a limiting value which is independent on the dissipation strength.~For increasing driving frequency, in the weak coupling regime (see Fig.~\ref{fig:dynboundcomp}, panel $\mathbf{b}$) $\mathcal{V}_{\rm{\scriptscriptstyle {TUR,ME}}}(\omega)$ exhibits two distinct peaks for $\omega$ slightly above and below resonance, and in the same region it also vanishes for two distinct frequency values.~The presence of a peak below resonance assures that a distinct frequency region exists where dynamic TUR violation occurs (see  Fig.~\ref{fig:dynboundcomp}, panel $\mathbf{b}$, and inset), ranging from frequencies $\omega\simeq 10^{-1}\Delta$ to the resonance region.~However, near resonance the dynamic TUR can be satisfied due to the vanishing of $\mathcal{V}_{\rm{\scriptscriptstyle {TUR,ME}}}(\omega)$.~For increasing dissipation strength, our numerical results show that the double-peak structure above and below resonance of $\mathcal{V}_{\rm{\scriptscriptstyle {TUR,ME}}}(\omega)$ tends to smooth down.~As a consequence, the region where violation is present reduces, while the resonance moves towards lower frequencies. For $\alpha>0.10$ (see  Fig.~\ref{fig:dynboundcomp}, panel $\mathbf{c}$-$\mathbf{d}$, and insets), the dynamic TUR is thus obeyed in the low-frequency region.
         
    Above resonance, a very different scenario sets in: here a value of the driving frequency exists such that the dynamic bound vanishes for all the values of dissipation strength, so that the converter obeys dynamic TUR.~Nevertheless, at higher frequencies, due to the decreasing behavior of $\mathcal{Q}_{\rm{\scriptscriptstyle {ME}}}(\omega)$, an even more pronounced violation of dynamic TUR with respect to the previous frequency range occurs.~As the dissipation strength increases, it can be noticed that $\mathcal{Q}_{\rm{\scriptscriptstyle {ME}}}(\omega)$ exhibits a slower decrease, and $\mathcal{V}_{\rm{\scriptscriptstyle {TUR,ME}}}(\omega)$ tends to a progressively lower limit (see Fig.~\ref{fig:dynboundcomp}, panels $\mathbf{c}$-$\mathbf{d}$),~\ie~the two curves become closer and the frequency region where dynamic TUR violation is present reduces.
          
    Our analysis shows that the frequency ranges where a violation of TUR occurs progressively narrow with increasing dissipation strength, both in the static and dynamic setting.~It thus shows that quantum coherence is the cause of the observed violations.~A further useful insight can be obtained by studying the behavior of  $\mathcal{Q}_{\rm{\scriptscriptstyle {ME}}}(\omega)$ and $\mathcal{V}_{\rm{\scriptscriptstyle {TUR,ME}}}(\omega)$ in the peculiar case of $\alpha=1/2$, where the converter shows a completely incoherent dynamics due to the influence of the bath (see Sec.~\ref{subsec:Toulouse}).
   
   \subsection{Performance at $\alpha=1/2$ in the scaling limit}\label{subsec:Toulouse}
   
   Below, we study the converter perfomance and the validity of dynamic TUR in the paradigmatic case of $\alpha=1/2$, the so-called Toulouse limit in the SBM literature \cite{Leggett,Weiss:open-quantum2}.~In the scaling limit $\omega_c\to\infty$, the path-integral expressions for the TLS magnetization and the two-time correlation functions, as well as the full moment generating functions of the energy exchange statistics for the driven SBM model in Eq.~\eqref{eq:TotalHamchap3} can be analytically solved \cite{SassettiWeiss:energyexchange}.~Besides of providing exact analitycal expressions for the Onsager matrix in Eq.~\eqref{eq:LinOnsager}, this strong coupling limit is highly interesting, as in the absence of external fields, the TLS system exhibits completely incoherent tunnelling dynamics.~We start by writing the solution for the magnetization along $z$ of the generic driven system as in Eq.~\eqref{eq:TotalHamchap3} in the limit $\alpha=1/2$, which reads \cite{SassettiWeiss:energyexchange}
    \begin{multline}\label{eq:Toul1}
    \ev{\sigma\ped{z}(t)}= e^{-\gamma t} + \Delta^2\int\limits_0^t \mathrm{d}\tau \int\limits_0^{t-\tau} \mathrm{d}s e^{-\gamma(s+\tau/2)} e^{-W(\tau)}\cdot\\
         \cdot\sin[G(t-s,t-s-\tau)],    
    \end{multline}
    where $\gamma=\Delta\ped{eff}(\alpha=\frac{1}{2})=\frac{\pi\Delta^2}{2\omega\ped{c}}$,~\ie~the Kondo frequency.~Here the functions $G(t),W(\tau)$ depend on the driving fields and on the properties of the bath respectively and read 
    \begin{equation}\label{eq:Gfunc}
    \begin{gathered}
    G(t_2,t_1)=\sum_{i=1,2}\int\limits_{t_1}^{t_2}\epsilon_i(t^{\prime})\mathrm{d} t^{\prime},\\
    W(\tau)=\log\left[\frac{\beta \omega\ped{c}}{\pi} \sinh\left[\frac{\pi \abs*{\tau}}{\beta}\right]\right].
    \end{gathered}
    \end{equation}
    Our interest is focused on the stationary value of the powers in Eq.~\eqref{eq:powerst} in the long-time limit,~\ie~$t\to\infty$, and in the linear response regime.~Hence, taking the mean over the driving period, we can rewrite the mean stationary powers as follows
    \begin{multline}\label{eq:statPowonehalf}
    \overline{P}_j =\frac{\Delta^2}{2}\frac{1}{\mathcal{T}}\int\limits_{0}^{\mathcal{T}}\mathrm{d}t\dot{\epsilon}_j (t)\int\limits_0^{+\infty} \mathrm{d}\tau \int\limits_0^{+\infty} \mathrm{d}s e^{-\gamma(s+\tau/2) -W(\tau)} \cdot\\ \cdot G(t-s,t-s-\tau).
   \end{multline}
   
    \begin{figure}[h!]
   	\centering 
   	\includegraphics[width=8.4 cm]{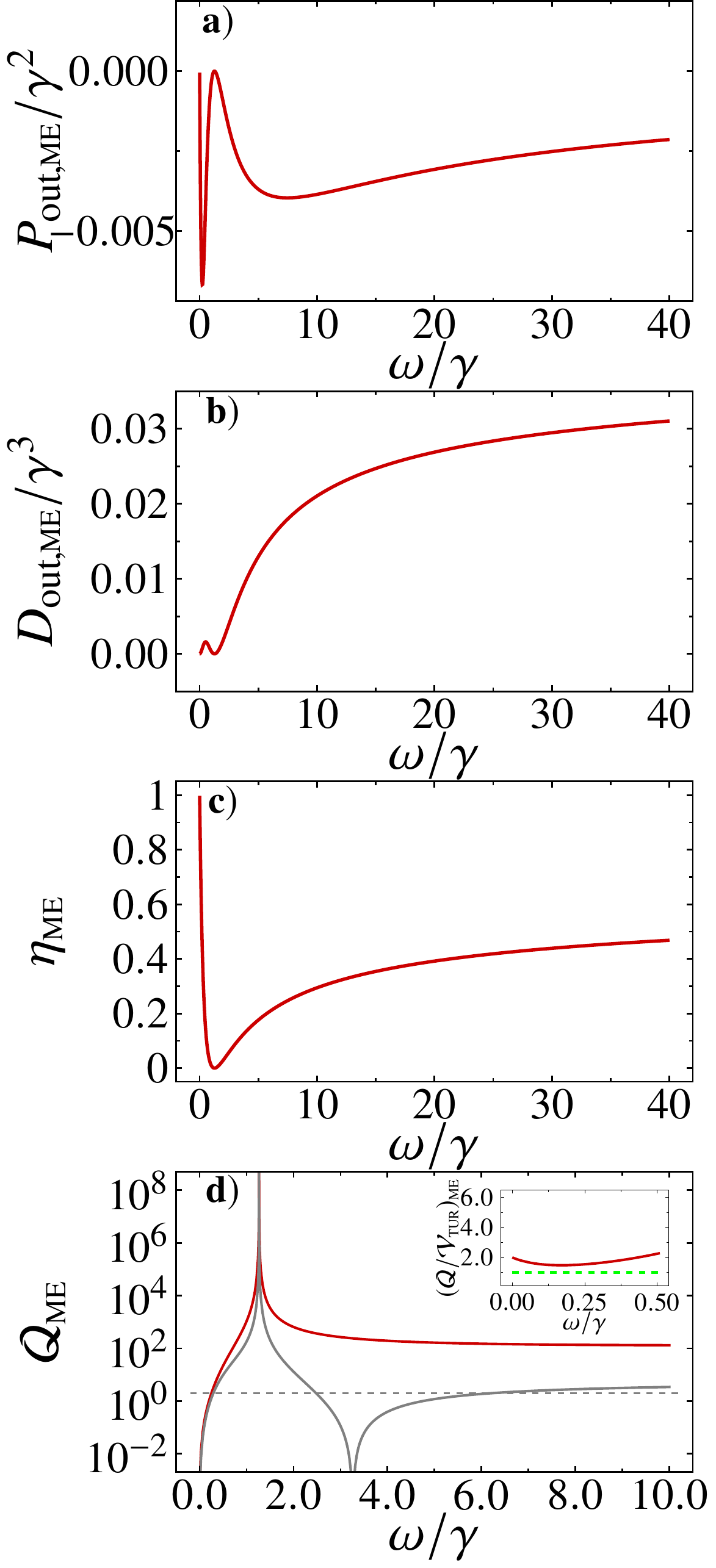}
   	\caption{Performance of the converter at ME, in the scaling limit, for coupling strength $\alpha=0.50$, inverse temperature $\beta=10\gamma^{-1}$, and $\epsilon_{2}=0.50\gamma$.~In panels $\mathbf{a}$-$\mathbf{c}$, the output power, the mean fluctuations, the efficiency curve are plotted versus the driving frequency $\omega$ respectively.~In panel $\mathbf{d}$, semilogarithmic plot of the tradeoff parameter $\mathcal{Q}_{\rm{\scriptscriptstyle {ME}}}(\omega)$ against  static and dynamic TUR bounds is shown.~Differently from Fig.~\ref{fig:Performance}, panel $\mathbf{b}$, in the high frequency regime the mean fluctuations (panel $\mathbf{b}$) grow monotonically with frequency.~As a consequence, the tradeoff parameter (panel $\mathbf{d}$ and inset) is above $\mathcal{V}_{\rm{\scriptscriptstyle {TUR,ME}}}(\omega)$,~\ie~the dynamic TUR cannot be violated in this regime.}
   	\label{fig:alpha05} 
   \end{figure}

    Notice that the nonlinear function of the driving fields in Eq.~\eqref{eq:Gfunc} has been replaced with its argument, which involves nothing more than integrals of the driving fields.~After several straightforward manipulations, by comparing Eq.~\eqref{eq:statPowonehalf} with Eq.~\eqref{eq:LinOnsager} the Onsager functions can be written as 
    \begin{equation}\label{eq:Onsalphaonehalf}
    \begin{gathered}
    \mathcal{L}_{11}(\omega)=\mathcal{L}_{22}(\omega)=\frac{1}{4(\omega^2 + \gamma^2)}\Big(\omega R_1(\omega) + \gamma R_2(\omega)\Big), \\
    \mathcal{L}_{12}(\omega)=\frac{1}{4(\omega^2 + \gamma^2)}\Big[\sin\varphi\Big(\omega R_1(\omega)+\gamma R_2(\omega)\Big) \\+ \cos\varphi\Big(\omega R_2(\omega)-\gamma R_1(\omega)\Big) \Big], \\
    \mathcal{L}_{21}(\omega,\varphi)=-\mathcal{L}_{12}(\omega,-\varphi), 
    \end{gathered}
    \end{equation}
    where the functions $R_1(\omega),R_2(\omega)$ can be recasted in terms of linear combinations of polygamma functions $\psi^{(m)}(z)$ with $m=0$ as follows
    \begin{equation}\label{eq:Ploygamma}
    \begin{gathered}
    R_1(\omega)=\Delta^2\int\limits_{0}^{+\infty}\mathrm{d}\tau e^{-\gamma\tau/2-W(\tau)}  \sin\omega\tau
    =\frac{i\gamma}{\pi}(\psi^{(0)}(z)-\psi^{(0)}(\overline{z})),\\
    R_2(\omega)=\Delta^2\int\limits_{0}^{+\infty}\mathrm{d}\tau e^{-\gamma\tau/2 -W(\tau)}  (1-\cos\omega\tau)=\\=\frac{\gamma}{\pi}\left(\psi^{(0)}(z) + \psi^{(0)}(\overline{z}) -2\psi^{(0)}(z^{\prime})\right), 
    \end{gathered}
    \end{equation}
    with $z=1/2+\gamma\beta/4\pi-i\omega\beta/2\pi\mbox{, }z^{\prime}=1/2+\gamma\beta/4\pi$, and  $\overline{z}$ is the complex conjugate of $z$.~In Fig.~\ref{fig:alpha05}, we plot the output power (panel $\mathbf{a}$), the output power fluctuations (panel $\mathbf{b}$), the efficiency (panel $\mathbf{c}$) and the tradeoff parameter (see Eq.~\eqref{eq:perfme} and the left hand side of Eq.~\eqref{eq:tur}) against the dynamic bound (panel $\mathbf{d}$), at ME, as a function of the driving frequency $\omega$, for a fixed value of the inverse temperature $\beta=10\Delta$.~As expected, the analytical results show that the converter performance noticeably degrades as compared with lower coupling strengths, being signaled by the high-frequency value of the efficiency curve.~However, it can also be noticed that in the same limit the power fluctuations take a different behavior with respect to Fig.~\ref{fig:Performance}, increasing with driving frequency as $D_{\rm{out,\scriptscriptstyle {ME}}}\simeq\epsilon_2^2 (\gamma/4)(1-\pi g^{-1} +(\pi^2/2)g^{-2})$ where $g=\abs{\log(\beta \omega/2 \pi)-\psi^{(0)}(z^{\prime})}$.~As a consequence, fluctuations increase with the driving frequency: above resonance, the tradeoff parameter $\mathcal{Q}_{\rm{\scriptscriptstyle {ME}}}$ reaches a minimum and then starts to increase as a function of $\omega$, while the dynamic bound drops to zero and subsequently slowly increases.~As a result, for $\alpha=1/2$, in all the investigated frequency region no violations of dynamic TUR can occur.~The absence of quantum coherence in the dynamics of the TLS working medium is the cause of the observed effect.   
        
    \section{Conclusions}\label{sec:conclusions}
    In this work, we analyzed the performance of an isothermal steady-state work-to-work quantum converter, taking as a working medium the simplest yet nontrivial quantum system,~\ie~a periodically-driven TLS, in the absence of TR symmetry, that is also permanently in contact with the thermal reservoir.~We studied the system by means of the numerically exact SIL method: we first established suitable model parameter ranges where work-to work conversion takes place; then, restricting to the linear response regime, we computed the output power, the fluctuations at ME as function of the driving frequency, for different values of dissipation strength.~From these results, we found evidence of a violation of the static and dynamic classical Markovian TUR, occurring in a wide range of model parameters.~Combining our numerical approach with known analytic results, we linked the observed violation to the degree of quantum coherence in the dynamics.~Although the converter dynamics is highly non-Markovian in the transient time, at least in the linear response regime and in the frequency range considered, we found the non-equilibrium stationary state to be only slightly affected by non-Markovian effects, suggesting that the influence of non-Markovianity in the observed TUR violation is limited.~In the near future, we plan to extend our analysis to the nonlinear response regime, with the aim to assess the converter performance under sufficiently high-intensity driving fields and understand if TUR violations are still present.~Another interesting extension of our work could be the study of the work-to-work converter performance under fast-forward protocols \cite{SelsE3909,Claeys:FloqEng,Villazon:fastforwardheat}, which have been recently proved to minimize irreversible loss in single and many-body quantum systems. 
          
  	\begin{acknowledgments}
  		We acknowledge the CINECA award under the ISCRA initiative (project QADS), for the availability of high-performance computing resources and support.
  	\end{acknowledgments}
	\appendix
	
    \section{Numerical approach}\label{app:SILapp}
    
    The nonequilibrium dynamics governed by Eq.~\eqref{eq:TotalHamchap3} can be studied numerically by evaluating the evolution operator $U(t, t_{0})$ of the whole TLS+bath system.~We simulate the unitary dynamics of the global state $\ket{\Psi(t)}$ of the system, so that the expectation value of any observable related to the TLS and the bath can be computed, along with two-time correlators. This task can be carried out by means of the following procedure: employ a discretization of the bath modes entering in Eq.~\eqref{eq:BathHamChap3}; then find a suitable truncation scheme of the bath Hilbert space, and apply the SIL method \cite{Cangemi:SIL,Lanczos1,Lanczos2}.~We discretize the bath modes by choosing a density of states $\rho(\omega)$ and by fixing the total number of bosonic modes $M\ped{mod}$ in the range $[0,2\omega\ped{c}]$.~We employ an exponentially decreasing density of states with frequency cutoff $\omega\ped{c}$
    \begin{equation}
    \rho(\omega)\propto\exp(-\frac{\omega}{\omega\ped{c}})\mbox{ , } \int\limits\ped{0}^{\omega\ped{c}}\rho(\omega)\mathrm{d}\omega=M\ped{mod}.
    \end{equation}    
    For each mode of frequency $\omega_k$, we choose the coupling strength $g(\omega_k)$ such that 
    \begin{equation}
    \rho(\omega_k)g^2(\omega_k)=2 \alpha \frac{\omega_k^{s}}{\omega\ped{c}^{s-1}}\eu^{-\frac{\omega_k}{\omega\ped{c}}}.
    \end{equation}    
    Our discrete system can mimick the theoretical model of a continuum set of modes,~\ie a thermal bath, as long as $M\ped{mod}$ is sufficiently high.~Every bath state is described by a set of basis states $\Set{\ket{n_{1},n_{2},\dots,n_{M\ped{mod}}}} $, where $n_{k}$ is the occupation number of the $k$-th bosonic mode of the bath.~In order to achieve a reliable truncation of the bath Hilbert space, we fix the absolute maximum number of bosonic excitations $N\ped{ph}$ with respect to the thermal equilibrium, and we restrict the description only to states for which $ \Delta n_{k}=n_{k}-n_{k}\api{eq} = \Set{0, \pm 1, \pm 2,\dots, \pm N\ped{ph}}$, with $\sum_k \abs{\Delta n_{k}}\leq N\ped{ph}$, where $n_{k}\api{eq}$ is the occupation number of the $k$-th bosonic mode at equilibrium.~It follows that this numerical approach can give an exact description of the physics up to terms in $\alpha^{N\ped{ph}}$.~The resulting dimension of the truncated Hilbert space of the TLS+bath system can be easily computed as   
    \begin{equation}
    \mathcal{N}=2\sum_{j=1}^{N\ped{ph}}\binom{N\ped{ph}}{j}\binom{M\ped{mod}}{j}.
    \end{equation} 
    Once the set of basis states has been fixed, we compute iteratively the quantum state of the system $\ket{\Psi(t)}$ at each time $t$.~By employing a discretization of the total evolution time interval in steps $\mathrm{d}t$, we perform a projection of the Hamiltonian evaluated at midpoint in each time interval $\rng{t}{t+\mathrm{d} t}$ into the $ n $-dimensional subspace $\mathcal{K}=\Set{\ket{\Psi(t)}, \ham\ket{\Psi(t)}, \dots, \ham^{n}\ket{\Psi(t)}}$ spanned by the Krylov orthonormal vectors $\Set{\ket{\Phi_{k}}}_{k=1}^{n}$, which can be computed using recursive Gram-Schmidt orthogonalization techniques.~The reduced Hamiltonian $\tilde{\ham}(t+\mathrm{d} t/2)=P \ham(t+\dd{t}/2) P^{\dagger}$, where $P$ is the projection operator in the subspace $\mathcal{K}$, can be easily diagonalized. The evolution operator in terms of the eigenstates of $\tilde{\ham}(t+\mathrm{d} t/2)$ reads       
    \beq\label{eq:projectedEv}
    \tilde{U}(t+\mathrm{d}t,t)\simeq \exp[-\iu \tilde{\ham}(t+\mathrm{d}t/2) \mathrm{d}t].
    \eneq   
    Eventually, we expand the state at previous time $t$ $\ket{\Psi(t)}$ in terms of the eigenvectors of $\tilde{\ham}\qty(t+\mathrm{d}t/2)$, and using \eqref{eq:projectedEv} we are able to compute the state at the end of the time interval $\ket{\Psi(t+\mathrm{d}t)}$ by means of matrix products.~The computation of the full ket state allows us to derive the density matrix of the system, from which we can numerically trace over the bath degrees of freedom and compute the reduced density matrix of the TLS.~The computation of two-time correlation functions of the kind $\ev{A(t)B(t^{\prime})}$ can be also achieved, noticing that the correlator can be written in the Heisenberg representation as follows
    \begin{equation}
    \ev{A(t)B(t^{\prime})}=\bra{\psi(0)}U^{\dagger}(t,0)AS(t,t^{\prime})BU(t^{\prime},0)\ket{\psi(0)},
    \end{equation}
    where $\ket{\psi(0)}$ is the initial state of the system and $S(t,t^{\prime})=U(t,0)U^{\dagger}(t^{\prime},0)$ is the $S$ matrix written in the Heisenberg representation, which admit the formal solution 
    \begin{equation}
    S(t,t^{\prime})=\hat{\mathcal{T}}\exp\qty[-i \int\limits_{t^{\prime}}^{t}H(\tilde{t})\mathrm{d} \tilde{t}].   
	\end{equation}
	It is thus evident that for $t^{\prime}<t$, the two-time correlator can be computed at the cost of doubling the Lanczos iteration, and eventually computing numerically the inner product of two different quantum states at time $t$.~Higher order correlators could also be computed with greater computation time.
	\section{Performance in the weak coupling regime}\label{app:weakcoup}
	In the weak couping limit, analytical closed expressions for the Onsager matrix in the absence of external fields, defined in Eq.\eqref{eq:Onsagercoeff} can be determined.~We start from the expression of the correlation function $C(\tau)$. In the limit of long times, it can be computed as in the following
	\begin{equation}
	C(\tau)=\Tr\left[\rho(\infty) \sigma_z(\tau)\sigma_z(0)\right],
	\end{equation}
     where $\rho(\infty)$ is the reduced density matrix of the system computed at sufficiently long times.~Following the weak damping limit, formulated in several works \cite{Carrega15:heatexchange,Grifo:epjb}, the correlation function takes the following form
     \begin{equation}\label{eq:weakdamp}
     C(\tau)= (A_1 \cos(\Omega t) + A_2\sin(\Omega t))e^{-\tilde{\gamma} t}, 
     \end{equation}   
     where $\Omega=\Delta\ped{eff}=\Delta (\Delta/\omega\ped{c})^{\alpha/(1-\alpha)}\left[\Gamma(1-2\alpha) \cos\pi \alpha\right]^{1/(2(1-\alpha))}$ is the effective oscillation frequency of the TLS, $\tilde{\gamma}=\pi\alpha\Omega/2 \coth(\beta \Omega/2 )$, $A_1=1$ and $A_2=\tilde{\gamma}/\Omega -i (\Omega/\Delta)^2 \tanh(\beta \Omega/2)$. The form \eqref{eq:weakdamp} is valid for $\alpha\ll 1$ (weak damping) and $\beta\Omega>1$. It thus follows that the elements of the Onsager matrix can be easily written as
     \begin{equation}\label{eq:Onsweak}
     \begin{gathered}
     \mathcal{L}_{11}(\omega)=\frac{\omega}{8}\left(\frac{\Omega}{\Delta}\right)^2 \left(\frac{4\Omega \omega \tanh(\frac{\beta \Omega}{2})\tilde{\gamma}}{(\tilde{\gamma}^2+(\Omega -\omega)^2)(\tilde{\gamma}^2+(\Omega +\omega)^2)}\right), \\
     \mathcal{L}_{12}(\omega,\varphi)=\frac{\omega}{8}\left(\frac{\Omega}{\Delta}\right)^2 \left(\frac{\tanh(\frac{\beta \Omega}{2})(\sin\varphi f(\omega) -\cos\varphi g(\omega))}{(\tilde{\gamma}^2+(\Omega -\omega)^2)(\tilde{\gamma}^2+(\Omega +\omega)^2)}\right),\\
     \mathcal{L}_{21}(\omega,\varphi)=-\mathcal{L}_{12}(\omega,-\varphi),\\
     \mathcal{L}_{22}(\omega)=\mathcal{L}_{11}(\omega),
     \end{gathered}
     \end{equation}
     where $f(\omega)= 4 \Omega \omega \tilde{\gamma}$ and $g(\omega)=2\Omega(\tilde{\gamma}^2 + \Omega^2 -\omega^2)$.
     
     For the sake of simplicity, we put $\varphi=0$.~Using Eq.~\eqref{eq:eff},~\eqref{eq:maxeffline}, the expression $1-\eta\ped{\scriptsize{ME}}$ can be rewritten in terms of the single parameter $z(\omega)=\mathcal{L}_{12}(\omega)/\mathcal{L}_{11}(\omega)$ as follows
     \begin{equation}\label{eq:efff}
     1-\eta\ped{\scriptsize{ME}}=2\frac{(\sqrt{1+z^2(\omega)} -1)}{z^2(\omega)}.
     \end{equation}
     From now on, we drop the subscript $\text{ME}$.~Eq.~\eqref{eq:efff} can be inverted to give $z(\eta)=2\sqrt{\eta}/(1-\eta)$.~Furthermore, from \eqref{eq:Onsweak} we can find that $z(\omega)=-(\tilde{\gamma}^2+\Omega^2 -\omega^2)/2\tilde{\gamma}\omega$, so that $z$ vanishes exacty at $\tilde{\omega}=\sqrt{\tilde{\gamma}^2 +\Omega^2}$.~We can thus obtain the two branches of the curve $\omega(\eta)$.~We focus on the positive branch,~\ie~$\omega>\tilde{\omega}$.~In the limit of high frequencies, we can thus write $\omega\simeq 4\tilde{\gamma} \sqrt{\eta}/(1-\eta)$.~At maximum efficiency, $P\ped{out},D\ped{out}$ read respectively     
     \begin{equation}\label{eq:powdout}
     \begin{gathered}   
     P\ped{out}={\epsilon_2}^2 \mathcal{L}_{11}(\omega)\frac{2(1+z^2) -z^2\sqrt{1+z^2} -2\sqrt{1+z^2}}{z^2},\\
     D\ped{out}={\epsilon_2}^2\frac{(1-\sqrt{1+z^2})^2}{z^2}\omega \coth(\frac{\beta \omega}{2})\mathcal{L}_{11}(\omega). 
     \end{gathered}
     \end{equation}         
     Inserting the expression of $\omega(\eta)\mbox{, }z(\eta)$, in the limit of high frequency,~\ie ~$\eta\to 1$, we find the leading contributions to Eq.~\eqref{eq:powdout} to be linear in $1-\eta$      
     \begin{equation}\label{eq:linslope}
     \begin{gathered}   
     P\ped{out}\simeq-\frac{{\epsilon_2}^2}{16}\left(\frac{\Omega}{\Delta}\right)^2\tanh(\frac{\beta \Omega}{2})\frac{\Omega}{\tilde{\gamma}}(1-\eta),\\
     D\ped{out}\simeq \frac{{\epsilon_2}^2}{8}\Omega \left(\frac{\Omega}{\Delta}\right)^2\tanh(\frac{\beta \Omega}{2})(1-\eta).
     \end{gathered}
     \end{equation}     
     A quantitative agreement of the previous expressions with SIL results reported in Fig.~\ref{fig:powerlaw} can be obtained, for sufficiently small coupling strengths $\alpha<0.001$. We also stress that, in the weak coupling regime, similar results could be achieved with a fully Markovian approach, based on Lindblad QME. 
\bibliography{paper}   
         
\end{document}